%% file: zeroaut-final.tex
\documentclass[11pt,a4paper]{article}
\usepackage{amsfonts,amsmath,amsthm,mathbbol,epsf,graphics,verbatim,amssymb,amscd,eucal,bbm}

\newcommand{\bdl}[2]
{\mbox{$
\renewcommand{\arraystretch}{0.5}
\begin{array}{c}{\scriptstyle #1}\\ {\scriptstyle #2} \end{array}
\renewcommand{\arraystretch}{2}
       $}}

\newtheorem{theorem}{Theorem}[section]

\newtheorem{proposition}[theorem]{Proposition}
\newtheorem{lemma}[theorem]{Lemma}

\newtheorem{definition}[theorem]{Definition}

\theoremstyle{remark}
\newtheorem{remark}[theorem]{Remark}

\def\Blackboardfont{\mathbb}

\newcommand{\moins}{ {\setminus} }

\newcommand{\pres}[2]{\langle \: #1 \mid #2 \: \rangle}

\def\rig{\text{Next}}
\def\Noact{\text{Noact}}
\def\lef{\text{Next}}
\def\leftt{\text{Left}}
\def\rightt{\text{Right}}

\def\cb{\cB}
\def\closcb{\bar{\cB}}

\def\B{{\Blackboardfont B}}

\def\F{{\Blackboardfont F}}

\def\Z{{\Blackboardfont Z}}

\def\N{{\Blackboardfont N}}
\def\R{{\Blackboardfont R}}

\def\un{{\mathbb 1}}

\def\cB{{\mathcal B}}

\def\cX{{\mathcal X}}

\def\iff{\Longleftrightarrow}
\def\eref#1{(\ref{#1})}

\def\corresp{\twoheadrightarrow}

\begin{document}
\sloppy

\title{\bf Zero-Automatic Queues and Product Form}

\author{Thu-Ha {\sc Dao-Thi} and  Jean {\sc Mairesse}
\thanks{LIAFA, CNRS-Universit\'e Paris 7, case
    7014, 2, place Jussieu, 75251 Paris Cedex 05, France. E-mail: {\tt
      (daothi,mairesse)@liafa.jussieu.fr}}}

\maketitle


\begin{abstract}
We introduce and study a new model: {\em 0-automatic
  queues}. Roughly, 0-automatic queues are
characterized by a special buffering mechanism evolving like a
random walk on some infinite group or monoid. The salient result
is that all stable 0-automatic queues have a product form stationary
  distribution and a Poisson output process. When
considering the two simplest and extremal cases of 0-automatic
queues, we recover the simple \emph{M/M/1} queue, and Gelenbe's
\emph{G}-queue with positive and negative customers.
\end{abstract}

\textsl{Keywords:} Queueing theory, M/M/1 queue, G-queue,
quasi-reversibility, product form, Quasi-Birth-and-Death process.

\smallskip

\textsl{AMS classification (2000):} Primary 60K25, 68M20.





\section{Introduction}

\begin{figure}[ht]
\[ \epsfxsize=300pt \epsfbox{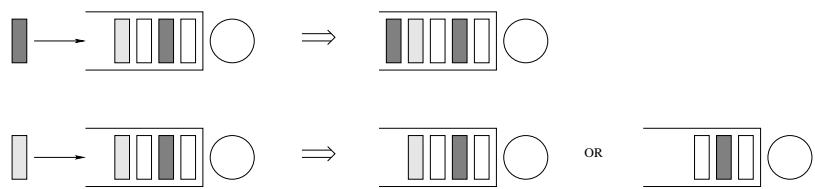} \]
\caption{A 0-automatic queue.}
\label{fi-0aut}
\end{figure}

Here is an informal description of a special type of 0-automatic
queue (corresponding to a free product of three finite
monoids). Consider a queue with a single server and an infinite
capacity buffer.
Customers are colored either in Red, Blue, or
Green, with a finite set of possible shades within each color:
$\Sigma_R, \Sigma_B, \Sigma_G$.
In the buffer, two consecutive customers of the
same color either cancel each other or merge to give a new
customer of the same color. Customers of different colors do not
interact.
This is illustrated in Figure \ref{fi-0aut}.

The shades get modified in the merging procedure, according
to an internal law: $\Sigma_i \times \Sigma_i \rightarrow \Sigma_i\cup
\{1\}$, with $1$ coding for the cancellation.
The only but crucial restriction is that each internal law should
be associative.

\medskip

We now give a more detailed account of the model and
results. Zero-automatic queues may be viewed as the synthesis of a
simple queue and a random walk on a 0-automatic pair. We first
recall these last two models.

The $M/M/1/\infty$ FIFO queue, or simply $M/M/1$ queue, is the
Markovian queue with arrivals and services occurring at constant
rate, say $\lambda$ and $\mu$, a single server, an infinite
capacity buffer, and a First-In-First-Out discipline. This is
arguably the simplest and also the most studied model in queueing
theory, with at least one book devoted to it~\cite{cohe82}. The
queue-length process is a continuous time jump Markov process and
its infinitesimal generator $Q$ is given by: $\forall n\in \N, \
Q(n,n+1)=\lambda, \ Q(n+1,n)=\mu$. Under the stability condition
$\lambda<\mu$, the queue-length process is ergodic, and its
stationary distribution $\pi$ is given by:
\begin{equation}\label{eq-mm1}
\pi(n) = \bigl(1-\frac{\lambda}{\mu} \bigr) \Bigl(\frac{\lambda}{\mu}\Bigr)^n \:.
\end{equation}
Besides, and this constitutes the celebrated Burke Theorem, the
departure process in equilibrium has the same law as the arrival
process.

\medskip

Let us introduce the a priori completely unrelated model of random
walk on a plain group studied in \cite{mair04,MaMa}.

Let $X$ be an infinite group or monoid with a finite set of
generators $\Sigma$. Let $\nu$ be a probability measure on $\Sigma$
and let $(x_i)_{i\in
  \N}$ be a sequence of $\Sigma$-valued i.i.d. r.v.'s of law $\nu$.
Let $(X_n)_n$ be the sequence of $X$-valued r.v.'s defined by:
$X_0=1_X, \ X_{n+1}= X_n\ast x_n = x_0\ast x_{1}\ast \cdots
\ast x_n$,
where $1_X$ is the unit element of $X$ and $\ast$ is the group or
monoid law. By definition, $(X_n)_n$ is a realization of the
random walk $(X,\nu)$.

We now assume that the pair $(X,\Sigma)$ is formed by a {\em plain} monoid
with {\em natural} generators. The
definition will be given in Section \ref{se-prel}. For the
moment, it suffices to say that the  elements of $X$ can be set in
bijection with a regular language $L(X,\Sigma)\subset \Sigma^*$.
The random walk $(X_n)_n$ is viewed as evolving on $L(X,\Sigma)$.
If $X_n=ua, a\in \Sigma$, and $x_n=b\in
\Sigma$, then
\begin{equation}\label{eq-buff}
X_{n+1} = u \ \text{ if } a\ast b =1_X, \quad X_{n+1} = uc \
\text{ if } a\ast b =c \in \Sigma, \quad X_{n+1} = uab \ \text{
otherwise} \:.
\end{equation}
Now assume that the random walk is transient. Let
$\nu^{\infty}(u\Sigma^{\N})$ be the probability that the random
walk goes to infinity in the ``direction'' $u$ (i.e.
$\nu^{\infty}(u\Sigma^{\N})=P\{ \exists N, \forall n\geq N, X_n
\in u\Sigma^*\}$). The following is the main result in
\cite{mair04}:
\begin{equation}\label{eq-rw}
\forall u=u_1\cdots u_n \in L(X,\Sigma), \quad \nu^{\infty}
(u\Sigma^{\N}) = \widehat{q}(u_1)\cdots
\widehat{q}(u_{n-1})\widehat{r}(u_n)\:,
\end{equation}
where $\forall a\in \Sigma, \ \widehat{q}(a)\in (0,1),
\widehat{r}(a)\in (0,1)$.

\medskip

The expressions in \eref{eq-mm1} and \eref{eq-rw} share a common
``multiplicative'' structure.
Guided by this analogy, we want to merge the two models together.
To that purpose, we make the following
elementary observation: if we block the server in an $M/M/1$ queue,
the number of waiting customers after $n$ arrivals is $A_n=n$. And
$(A_n)_n$ can be viewed as the (not so random) random walk on the
pair $((\N,+),\{1\})$ associated with the probability $\nu: \
\nu(1)=1$.

Now, replace the trivial random walk $(A_n)_n$ by another, more
complex, random walk $(X_n)_n$ on a plain triple $(X,\Sigma,\nu)$.
Hence, the random walk $(X_n)_n$ constitutes the buffering
mechanism in a queue with a blocked server. A {\em 0-automatic
queue} is the model obtained when unblocking the server. The set
$\Sigma$ is the set of possible {\em classes} for customers. Customers
arrive at constant rate $\lambda$. Upon
arrival, a new customer (class $b$) interacts with the customer
presently at the back-end of the buffer (class $a$), according to
\eref{eq-buff}. At the front-end of the buffer, customers are
served at constant rate $\mu$. 

\medskip

Let us comment on the name
{\em zero-automatic}. Plain groups, see \eref{eq-plainm}, are
{\em automatic} in the sense of Epstein et. al~\cite{ECHLPT}. Automatic
groups form an important class of groups extensively studied in
geometric group theory, the adjective ``automatic'' referring to the
existence of automata to recognize and multiply elements of the
group. Now the pairs $(X,\Sigma)$ formed by a plain group with natural
generators satisfy the {\em 0-fellow traveller property}, see
\cite{ECHLPT}. It was proposed in \cite{mair04} to call 
$(X,\Sigma)$ a {\em 0-automatic pair}. By extension, a queue built
upon $(X,\Sigma)$ is {\em 0-automatic}. The name is also supposed
to evoke the local aspect of the interactions between customers in the
buffer, see \eref{eq-buff}. 

\medskip

Let $\widehat{\gamma}$ be the drift or rate of escape to infinity of the random
walk $(X_n)_n$. We prove in Section \ref{se-stab} that the stability
condition for the 0-automatic
queue associated with $(X_n)_n$ is: $\lambda\widehat{\gamma} < \mu$. Under this
condition, we prove in Section \ref{se-main} that the stationary
distribution $p$ for the queue-content process has a
``multiplicative'' structure:
\begin{equation}\label{eq-0autq}
\forall u=u_1\cdots u_n \in L(X,\Sigma), \quad p(u) = (1-\rho)\rho^n
q(u_1)\cdots q(u_{n-1}) r(u_n) \:,
\end{equation}
for some numbers $\rho\in (0,1), \ \forall a\in \Sigma, \ q(a)\in
(0,1),r(a)\in (0,1)$. (These numbers are in general different from
their counterparts in \eref{eq-mm1} and \eref{eq-rw}.)
Furthermore, the departure process from the
queue is a Poisson process of rate $\rho\mu$. Thus we have an analog
of Burke Theorem for all 0-automatic queues. Using standard
terminology, 0-automatic queues are {\em quasi-reversible}.

To be more precise, given $(X_n)_n$, several variants of 0-automatic
queues can be defined depending on the way customers are incorporated
in an empty queue (boundary condition). There is
precisely {\em one} choice for
which the result in \eref{eq-0autq} holds. The numbers $\rho,
q(\cdot ), r(\cdot )$, as well as the right boundary condition,
are obtained implicitly via the unique
solution of a set of algebraic equations, see Theorems
\ref{th-main} and \ref{th-uniq} for a precise statement.

\medskip

Aside from the free monoid, the next simplest example of a
plain monoid is the free group over one generator:
$(\Z,+)$. The 0-automatic queues associated with $(\Z,\{-1,1\})$ 
are variations of Gelenbe's
G-queues, or queues with positive and negative customers, which
were quite extensively studied in the 90's, see \cite{gele91,FGSu}
and the bibliography in \cite{GePu}. General 0-automatic queues
can be viewed as a wide generalization of this setting. 
Indeed, in a 0-automatic queue, different types of tasks (customers) can
be modelled. Let us detail four of them which form a
representative sample, without exhausting all the types within the
realm of 0-automaticity.

\medskip

- Classical type. Tasks are processed one by one with no
  simplification occurring in the buffer: $aa=aa$. The corresponding
  pair is $(\N,\{1\})\sim (\{a\}^*,\{a\})$.

\medskip

- Positive/negative type. Tasks are either positive ($a$) or
negative
  ($a^{-1}$) and two consecutive tasks of opposite signs cancel each
  other: $aa^{-1}=a^{-1}a=1$. The corresponding pair is
  $(\Z,\{1,-1\})\sim (\F(a),\{a,a^{-1}\})$.  The relevance of this type
  for applications is discussed in \cite{GePu}.

\medskip

- ``One equals many'' type. It takes the same time to process one
or
  several consecutive instances of the same task: $aa=a$. Think for instance of a ticket
  reservation where the number of requests is only reflected by
  an integer value in a menu-bar choice. The corresponding pair is
  $(\B,\{a\})$ where $\B$ is the Boolean monoid $\B = \pres{a}{a^2=a}$.

\medskip

- ``Dating agency'' type. Two instances of the same task cancel
each
  other: $aa=1$. Think of a task as being a tennis player looking for
  a partner (to be provided by the server); when two such tasks
  are next to each other in the buffer, they leave to play a
  game instead of waiting in line. The corresponding pair is
  $(\Z/2\Z=\pres{a}{a^2=1},\{a\})$. Instead of tennis players, we may
  consider music trio players, bridge players, etc, the corresponding
  group being $\Z/3\Z, \Z/4\Z$, etc.

\medskip

To model a server where several of the above types (and possibly
several copies of the same type) can be processed, one just has to
perform the free product of the corresponding monoids or groups (see
Section \ref{sse-mg} for the definition). 

\medskip


The $M/M/1$ queue is the basic primitive for building Jackson
networks, which have the remarkable property of having a
``product-form'' stationary distribution. More generally, networks
made of quasi-reversible nodes tend to have a product form
distribution, see for instance \cite{serf99}. In a subsequent
work~\cite{DaMa06}, we prove that it is indeed the case for Jackson-type
and Kelly-type
networks of 0-automatic queues.

\medskip

A preliminary version without proofs of the present paper has appeared
in the conference proceedings~\cite{DaMa05}. 

\section{Preliminaries}\label{se-prel}

{\em Notations.} We denote respectively by $\Z$, $\N$ and $\R_+$ the
integers, nonnegative integers and reals. We 
denote by $\N^*$ and $\R_+^*$ the positive integers and reals.
The symbol $\sqcup$ is used for the disjoint union of sets.
Given a set $T$ and $S\subset T$, define  $\un_S: T
\rightarrow \{0,1\}$ by $\un_S(u)=1$ if $u\in S$ and $\un_S(u)=0$ otherwise.
Given a set $T$, a
vector $x\in \R^T$, and $S\subset T$, set
$x(S)=\sum_{u\in S} x(u)$.

\medskip

Let us recall the needed material on random walks on plain monoids. 
The presentation follows \cite{mair04,MaMa}.

\subsection{Monoids and groups}\label{sse-mg}

Given a set $\Sigma$, the free monoid generated by $\Sigma$ is
denoted by $\Sigma^{*}$. The unit element is denoted by $1$ or $1_{\Sigma^*}$. As
usual, the elements of $\Sigma$ and $\Sigma^{*}$ are called {\em
letters} and {\em words}, respectively. The subsets of
$\Sigma^{*}$ are called {\em languages}. The {\em length} (number
of letters) of a word $u$ is denoted by $|u|_{\Sigma}$.

Let $(X,\ast)$ be a group or monoid with set of generators
$\Sigma$. The unit element of $X$ is denoted  by $1_{X}$. When $X$
is a group, the inverse of $x\in X$ is denoted by $x^{-1}$. We
always assume that: $1_X \not\in \Sigma$, and in the group case
that: $x\in \Sigma \implies x^{-1} \in \Sigma$. The {\em length}
with respect to $\Sigma$ of an element $x$ of $X$ is:
\begin{equation}
|x|_{\Sigma} = \min \{ k \mid x = a_{1}\ast \cdots \ast a_{k},
a_{i} \in \Sigma \}
\end{equation}

The {\em Cayley graph} $\cal X$$(X,\Sigma)$ of $X$ with respect to
$\Sigma$ is the directed graph with nodes $X$ and arcs
$u\rightarrow v$ if $\exists a\in \Sigma, \ u\ast a =v$.

\medskip

Consider a relation $R\subset \Sigma^*\times \Sigma^*$, and let
$\sim_R$ be the least congruence on $\Sigma^*$ such that $u \sim_R
v$ if $(u,v)\in R$. Let $X$ be isomorphic to the quotient monoid
$(\Sigma^* / \sim_R)$. We say that $\pres{\Sigma}{u=v, (u,v)\in
R}$ is a {\em
  monoid presentation} of $X$
and we write $X=\pres{\Sigma}{u=v, (u,v)\in  R}$.

Given a set $S$, denote by $\F(S)$ the free group generated (as a
group) by $S$. Let $S^{-1}$ be the set of inverses of the
generators. A monoid presentation of $\F(S)$ is
\begin{equation}\label{eq-freegroup}
\F(S) = \pres{S\sqcup S^{-1}}{aa^{-1}=1 ,
  a^{-1}a= 1, \ \forall a\in S}\:.
\end{equation}

Given two groups or monoids $X_1$ and $X_2$, we denote by
$X_1\star X_2$ the {\em free product} of $X_1$ and $X_2$. Roughly,
the elements of $X_1\star X_2$ are the finite alternate sequences
of elements of $X_1\moins\{1_{X_1}\}$ and $X_2\moins\{1_{X_2}\}$,
and the law is the concatenation with simplification. More
rigorously, the definition is as follows. Set $S = X_1\sqcup X_2$.
The {\em free  product} $X_1\star X_2$ is defined by the monoid
presentation:
\begin{equation*}\label{eq-freeproduct}
\pres{\ S \ } {\ (\forall u,v \in S^*, \forall i\in \{1,2\}), \
u1_{X_i}v = uv, (\forall a,b,c \in X_i, \text{ s.t. } c=a\ast b),
\ uabv = ucv \ } \:.
\end{equation*}
If $X_1$ and $X_2$ are groups, then $X_1\star X_2$ is also a
group. The free product of more than two groups or monoids is
defined analogously.

The Cayley graph of the group $\Z/2\Z\star \Z/3\Z$ is
represented on Figure \ref{fi-z2z3} (left). 

\subsection{Plain monoids and groups}\label{sse-pmg}

A {\em plain monoid} is a monoid $X$ of the form
\begin{equation}\label{eq-plainm}
X = S^* \star \F(T) \star X_1\star \cdots \star X_k \:,
\end{equation}
where $S$ and $T$ are finite sets and $X_1,\dots, X_k$ are finite
monoids. A {\em plain group} is a plain monoid which is also a
group. A plain monoid $X$ defined as in \eref{eq-plainm} is a
plain group iff $S=\emptyset$ and $X_1,\cdots, X_k$ are groups.

Define
\begin{equation}\label{eq-sigma}
\Sigma = S \sqcup T \sqcup T^{-1} \sqcup X_1\setminus\{1_{X_1}\}
\sqcup \cdots \sqcup X_k\setminus\{1_{X_k}\}\:.
\end{equation}
The set $\Sigma$ is a finite set of generators of $X$, that we
call {\em natural} generators. Define the language $L(X,\Sigma) \subset \Sigma^{*}$ by:
\begin{equation}\label{eq-loca}
L(X,\Sigma) = \bigl\{u_{1}\cdots u_{k} \mid \forall i< k, u_{i}
\ast u_{i+1} \notin \Sigma \cup \{1_{X}\} \bigr\}\:.
\end{equation}
It is easily seen that the set $L(X,\Sigma)$ is in bijection with the
group elements. Below we often identify $X$ and $L(X,\Sigma)$. 
The following is a consequence of the definition of a plain monoid~: 
\begin{equation}\label{eq-ref1}
a,b\in \Sigma, \qquad a\ast b \in \Sigma \cup \{1_X\} \ \iff \  b\ast a \in
\Sigma \cup \{1_X\} \:.
\end{equation}
To see that \eref{eq-ref1} holds, it is sufficient to check it case by
case. 
It is convenient to introduce the sets: $\forall a \in
\Sigma$,
\begin{equation}\label{eq-next}
\ \rig(a) = \{ b \in \Sigma \mid b \ast a \notin \Sigma \cup
\{1_{X}\}\} \nonumber= \{ b \in \Sigma \mid a \ast b \notin \Sigma
\cup \{1_{X}\}\}\:.
\end{equation}
Observe that $L(X,\Sigma) =  \bigl\{u_{1}\cdots u_{k} \mid \forall
i, \  u_{i-1} \in \lef( u_{i}) \bigr\} = \bigl\{u_{1}\cdots u_{k}
\mid \forall i, \  u_{i+1} \in \rig( u_{i}) \bigr\}$.

Next property, to be used later on, is another direct consequence of
the definition of a plain monoid~:
\begin{equation}\label{eq-ref2}
a\ast b \in \Sigma \ \implies \ \rig(a) = \rig(b) = \rig(a*b)\:.
\end{equation}

\medskip

Consider the directed {\em graph of successors} $(\Sigma,\rightarrow)$ where 
\begin{equation}\label{eq-graph}
a\rightarrow b \ \ \ \text{if} \ \ \ b\in \rig(a) \:.
\end{equation}
Except in the case $X=\F(T), |T|=1$, observe that the graph $(\Sigma,\rightarrow)$ is strongly
connected. 



\subsection{Random walks on monoids and groups}

Let $(X,\ast)$ be a group or monoid with finite set of generators
$\Sigma$. Let $\nu$ be a probability distribution over $\Sigma$.
Consider the Markov chain on the state space $X$ with one-step
transition probabilities given by: $\forall x \in X, \forall a \in
\Sigma$, $P_{x,x\ast a} = \nu(a)$. This Markov chain is called the
\textit{(right) random walk} (associated with) $(X,\nu)$.

Let $(x_{n})_n$ be a sequence of i.i.d. r.v's distributed
according to $\nu$. Set
\begin{equation}\label{7}
X_{0} = 1_{X}, \;\; \mbox{} X_{n+1} = X_{n}\ast x_{n}=x_{0}\ast
\cdots \ast x_{n} \:.
\end{equation}

Then $(X_n)_n$ is a realization of the random walk $(X,\nu)$. For
all $x, y \in X$, we have $|x\ast y |_{\Sigma} \leq
|x|_{\Sigma}+|y|_{\Sigma}$. Applying Kingman's Subadditive Ergodic
Theorem yields the following (first noticed by Guivarc'h
\cite{guiv80}): there exists $\gamma \in \mathbb{R}_{+}$ such that
\begin{equation}\label{eq-guiv}
\lim_{n \rightarrow \infty} \frac{|X_{n}|_{\Sigma}}{n} = \gamma
\;\; \mbox{ a.s and in $L^{p}$},
\end{equation}
for all $1\leq p <\infty$. We call $\gamma$ the {\em drift} of the
random walk.

\begin{figure}[ht]
\[  \input{z2z3.pstex_t} \]
\caption{The random walk $(\Z/2\Z \star \Z/3\Z, \nu)$.}
\label{fi-z2z3} 
\end{figure}
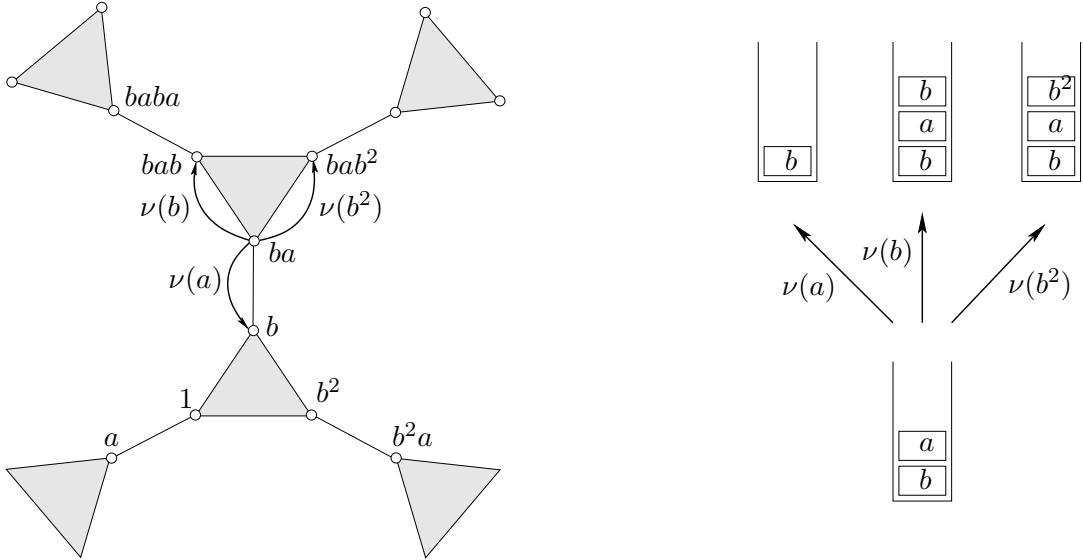

To illustrate, consider the plain group $X=\Z/2\Z \star \Z/3\Z =
\pres{a}{a^2=1}\star \pres{b}{b^3=1}$ and the natural generators
$\Sigma = \{a,b,b^2=b^{-1}\}$. Let $\nu$ be a probability measure on
$\Sigma$. 
On the left of Figure \ref{fi-z2z3}, we
have represented a finite part of the infinite Cayley graph $\cX(X,\Sigma)$,
and the one-step transitions of the random walk $(X,\nu)$ starting
from the state $ba$. On the right of the figure, 
we show the same one-step transitions on the group elements viewed as
words of $L(X,\Sigma)$ (written from bottom to top).

\subsection{Random walks on plain monoids and groups}

It is convenient to introduce the notion of a plain triple.

\begin{definition}\label{de-0aut3}
A triple $(X,\Sigma,\nu)$ is {\em plain} if: (i) $X$ is an
infinite plain monoid not isomorphic to $\Z$ or $\Z/2\Z \star
\Z/2\Z$; (ii) $\Sigma$ is a set of natural generators; (iii) $\nu$
is a probability measure whose support is included in $\Sigma$ and
generates $X$.
\end{definition}

\begin{proposition}\label{pr-transient}
If $(X,\Sigma, \nu)$ is a plain triple, then the random walk
$(X,\nu)$ is transient.
\end{proposition}

If $X$ is an infinite plain monoid with the support of $\nu$
generating $X$, there are only two cases in
which $(X,\nu)$ is not transient: (1) the triple $(\Z,\{-1,1\},
\{1/2,1/2\})$; (2) the triples $(\Z/2\Z\star \Z/2\Z ,\{a,b\},
\nu)$, for any $\nu$, where $a$ and $b$ are the respective
generators of the two cyclic groups. Since $\Z$ and $\Z/2\Z \star
\Z/2\Z$ have been excluded from consideration, then the random
walk $(X,\nu)$ is transient, see \cite{mair04} for details.

The case $X=\Z$ is specific. Some of the results below remain true
but not all of them. For simplicity, we treat this case separately in \S
\ref{se-examples}.

\medskip

Define:
\begin{equation}\label{eq-cb}
\cb = \{x\in \R^{\Sigma} \mid \forall i, x(i) >0, \ \sum_i x(i) =1
\}, \qquad \closcb = \{x\in \R^{\Sigma} \mid \forall i, x(i) \geq
0, \ \sum_i x(i) =1 \}\:.
\end{equation}

The Traffic Equations play an essential role in the study of the
random walk $(X,\nu)$.

\begin{definition}\label{de-TE}
The {\em Traffic Equations (TE)} associated with a plain triple
$(X,\Sigma,\nu)$ are the equations of the variables $(x(a))_{a\in
\Sigma}\in \R_+^{\Sigma}$ defined by: $\forall a\in \Sigma$,
\begin{equation}\label{eq-TE}
x(a)  =  \nu(a)x(\rig(a)) + \sum_{b\ast d = a} \nu(b)x(d)  +
\sum_{\bdl{d\in \lef(a)}{b\ast d=1_X}} \nu(b)
\frac{x(d)}{x(\rig(d))} x(a) \:.
\end{equation}
An {\em admissible solution} is a solution belonging to  $\cb$.
\end{definition}

By multiplying both sides of \eref{eq-TE} by $\prod_b x(\rig(b))$,
we obtain a new set of Equations without denominators. With some
abuse, a solution $r$ in $\closcb$ of this last set of Equations
is still called a solution of the TE.

Next result can be easily deduced from the proof of \cite[Theorem
  4.5]{mair04}.

\begin{proposition}\label{pr-solTE}
Let $(X,\Sigma,\nu)$ be a plain triple. The Traffic Equations have
a unique admissible solution.
\end{proposition}

The interest of Proposition \ref{pr-solTE} is that the harmonic
measure and the drift can be expressed as a function of the
solution to the TE.
Define the set
$L^{\infty}\subset \Sigma^{\N}$ by
\begin{equation}\label{eq-Linfty}
L^{\infty}= \{u_0u_1\cdots u_k \cdots  \in \Sigma^{\N} \mid \forall i
\in \N, u_{i+1}\in \rig(u_i) \}\:.
\end{equation}
A word belongs to $L^{\infty}$ iff all its finite prefixes belong to
$L(X,\Sigma)$. The set $L^{\infty}$ should be viewed as the
``boundary'' of $X$.

\medskip

Let $(X_n)_n$ be a
realization of the random walk which is transient by Proposition
\ref{pr-transient}. 
The {\em harmonic measure} of the random walk is the probability
measure $\nu^{\infty}$ on $L^{\infty}$ with finite-dimensional marginals defined by:
\[
\forall u_1\cdots u_k \in L(X,\Sigma), \
\nu^{\infty}(u_1\cdots u_k\Sigma^{\N}) = P \{ \exists N, \forall n\geq
N, \ X_n \in u_1\cdots u_k\Sigma^* \}\:.
\]
This defines indeed a measure on $L^{\infty}$ because the random walk
is transient, and because $X_n$ and $X_{n+1}$ differ by at
most their last symbol.
Intuitively, the harmonic measure $\nu^{\infty}$ gives the direction in which $(X_n)_n$
goes to infinity.

For a proof of next result, see \cite[Theorem 4.5]{mair04} and also
\cite[Theorem 3.3]{MaMa}. In
the specific case of the free group, the result appears in
\cite{DyMa,SaSt}, see also the survey \cite{ledr00}.

\begin{theorem}\label{th-rw}
Let $(X,\Sigma,\nu)$ be a plain triple. Let
$\widehat{r}=(\widehat{r}(a))_{a\in \Sigma}$ be the unique
  admissible solution to the Traffic Equations.
Set $\widehat{q}(a)= \widehat{r}(a)/\widehat{r}(\rig(a))$, for all
$a\in \Sigma$. The harmonic measure $\nu^{\infty}$ of the random
walk $(X,\nu)$ is given by:
\begin{equation}
\forall u_1\cdots u_k, \quad \nu^{\infty}(u_1 \cdots u_k\Sigma^{\N})
= \widehat{q}(u_1) \cdots \widehat{q}(u_{k-1}) \widehat{r}(u_k)\:.
\end{equation}
The drift of the random walk is given by:
\begin{equation}\label{eq-drift}
\widehat{\gamma} = \sum_{a\in \Sigma} \nu(a) \bigl[ \widehat{r}(\rig(a))  -
  \sum_{b \mid a\ast b =1_X} \widehat{r}(b) \bigr]\:.
\end{equation}
\end{theorem}

\section{The Zero-Automatic Queue}\label{se-0autq}

We first define the 0-automatic queue informally, before doing it
formally in Definition \ref{de-0aut}. 
Let $X$ be a plain monoid, $\Sigma$ be a set of natural
generators, and $\nu$ a probability measure on $\Sigma$. 
The associated
0-automatic queue is formed by a simple single server queue with
FIFO discipline and an infinite capacity buffer in which the
buffering occurs according to the random walk $(X,\nu)$. It is a
multi-class queue (classes $\Sigma$) but the class does not
influence the way customers get served, only the way they get
buffered.

\medskip

More precisely, the instants of customer arrivals are given by a
Poisson process of rate $\lambda$, and each customer carries a
mark, or {\em class}, which is an element of $\Sigma$. The
sequence of marks is i.i.d. of law $\nu$. Upon arrival, a new
customer interacts with the customer presently at the back-end of
the buffer, and depending on their respective classes, say $b$ and
$a$, one of three possible events occurs: (i) if $b\ast a =1_X$,
then the two customers leave the queue; (ii) if $b\ast a=c\in
\Sigma$, then the two customers merge to create a customer of type
$c$; (iii) otherwise, customer $b$ takes place at the back-end of
the buffer, behind customer $a$. In the mean time, at the
front-end of the buffer, the customers are served one by one and
at constant rate $\mu$ by the server. To be complete, one needs to
specify how customers are incorporated when the buffer is empty.
Several variants may be considered, and we view this ``boundary
condition'' as an additional parameter of the model.
The resulting flexibility in the definition of a 0-automatic queue
will turn out to be a crucial point.

\medskip

According to the above description, the queue-content (the
sequence of classes of customers in the buffer) is a continuous
time jump Markov process. The more formal definition of the queue
is given via the infinitesimal generator of this process.

\begin{definition}[Zero-automatic queue]\label{de-0aut}
Consider a plain triple $(X,\Sigma,\nu)$.
Let $L(X,\Sigma)$ be the
set of words defined in \eref{eq-loca}. Consider
$r \in \closcb$, see \eref{eq-cb},  and $\lambda, \mu \in \R_+^*$.
The {\em 0-automatic queue} of type $(X,\Sigma,\nu,r,\lambda,\mu)$
is defined as follows. The {\em
  queue-content} $(M(t))_{t\in \R_+}$ is a continuous time
jump Markov process on the state space $L(X,\Sigma)$ with infinitesimal generator
$Q$ defined by: $\forall u=u_n\cdots u_1 \in L(X,\Sigma) \setminus \cup_{a\in
  \Sigma}\{a\}^*$,
\begin{equation}\label{eq-main}
\left\{ \begin{array}{lcll}
Q(u,bu) & = & \lambda\nu(b)\ ,  & \forall b \in \lef(u_n)\\
Q(u,cu_{n-1}\cdots u_1) & = & \lambda \sum_{b \mid b\ast u_n=c}
\nu(b) \ ,&
\forall c \in \Sigma\setminus\{u_n\}, \ \exists b \in \Sigma, \ b\ast u_n=c \\
Q(u,u_{n-1}\cdots u_1)  & = & \lambda \sum_{b\mid b\ast u_n=1_X} \nu(b) & \\
Q(u,u_{n}\cdots u_2 )  & = & \mu &
\end{array} \right.
\end{equation}
and, for all $a\in \Sigma$ such that $a \in \rig(a)$, and for all
$n\geq 1$,
\begin{equation}\label{eq-main2}
\left\{ \begin{array}{lcll}
Q(a^n,ba^n) & = & \lambda\nu(b)\ ,  & \forall b \in \lef(a)\\
Q(a^n,ca^{n-1}) & = &  \lambda \sum_{b \mid b\ast a=c} \nu(b) \ ,&
\forall c \in \Sigma\setminus\{a\}, \ \exists b \in \Sigma, \ b\ast a=c \\
Q(a^n,a^{n-1})  & = & \mu + \lambda \sum_{b\mid b\ast a=1_X}
\nu(b)
\end{array} \right.
\end{equation}
and, finally, the boundary condition is,
\begin{equation}\label{eq-bound}
Q(1_{\Sigma^*}, a) \ = \ \lambda \nu(a) r(\rig(a))\ , \quad
\forall a \in \Sigma\:.
\end{equation}
We denote by $M/M/(X,\Sigma)$ any 0-automatic queue of type
$(X,\Sigma,\nu,r,\lambda,\mu)$.
\end{definition}

\remark\label{rm-bc} The intuition behind the form of the boundary
condition is as follows: the buffer-content is viewed as the
visible part of an iceberg consisting of an infinite word of
$L^{\infty}$, see \eref{eq-Linfty}. When the buffer is empty, new
customers are incorporated depending on the invisible part of the
iceberg, whose first marginal is assumed to be $r$. This last
point will find an a-posteriori justification in Theorem
\ref{th-main}.

\medskip

The simplest example of 0-automatic queue is the one associated
with the free monoid $(\N,+)$. The triple $(\Z,\{-1,1\},\nu)$,
where $\nu$ is a probability measure on $\{1,-1\}$, is not plain.
However, it is simple and interesting to generalize Definition \ref{de-0aut}
in order to define a 0-automatic queue associated with the free
group $(\Z,+)$. We now discuss the 0-automatic queues associated with 
$(\N,+)$ and $(\Z,+)$. 

\paragraph{The simple queue.}

Consider the free monoid $X=\{a\}^*=\{a^k,k\in \N\}$ over the single
generator set
$\Sigma=\{a\}$.
Hence, for any $\lambda, \mu \in \R_+^*$, there is only one
possible associated queue: $(X,\Sigma, \nu,r,\lambda,\mu)$, where
$\nu(a)=r(a)=1$. By specializing the infinitesimal generator $Q$
given in Definition \ref{de-0aut}, we get: $\forall n \in \N$,
\[
Q(a^n,a^{n+1}) = \lambda, \quad Q(a^{n+1},a^{n}) = \mu \:.
\]
This is the simple $M/M/1/\infty$ FIFO queue with arrival rate
$\lambda$ and service rate $\mu$.

\paragraph{The G-queue.}

Consider the free group $X=\F(a)=\{a^k,k\in \Z\}$ and the
set of generators $\Sigma=\{a, a^{-1}\}$. Let $\nu$ be a probability
measure on
$\Sigma$ such that $\nu(a)>0, \nu(a^{-1})>0$. Consider $r\in \cb$ and $\lambda,
\mu \in
\R_+^*$. The 0-automatic queue
$(\F(a),\Sigma,\nu,r,\lambda,\mu)$ has an infinitesimal
generator $Q$ given by: $\forall n \in \N$,
\begin{equation*}
\left\{ \begin{array}{lcllcl}
Q(a^n,a^{n+1}) & = &  \lambda \nu(a), & \ Q(a^{n+1}, a^n) & = &   \mu +
\lambda\nu(a^{-1}) \\
Q(a^{-n}, a^{-(n+1)}) & = &    \lambda\nu(a^{-1}),  &  \ Q(a^{-(n+1)},
a^{-n}) & = & \mu + \lambda\nu(a) \\
Q(1_{\Sigma^*},a) & = &   \lambda \nu(a)r(a), & \ Q(1_{\Sigma^*},a^{-1}) & = &
\lambda
\nu(a^{-1})r(a^{-1}) \:.
\end{array}\right.
\end{equation*}
This is close to the mechanism of the G-queue, a queue with positive and
negative customers introduced by Gelenbe \cite{gele91,GePu}.
With respect to the G-queue, one originality of the
$M/M/(\F(a),\Sigma)$ queue is that negative and
positive customers play
symmetrical roles. Another one is the treatment of the boundary
condition. 

\medskip

Since the triple $(\F(a),\{a^{-1},a\},\nu)$ is not plain according to
Def. \ref{de-0aut3}, the above queue is not covered by the results in
Sections \ref{se-stab} and \ref{se-main}. However, part of the results
remain true, and we come back specifically to this model in 
Section \ref{sse-G} and \ref{ssse-fg}. 




\paragraph{Extension.}
It is possible to generalize Definition \ref{de-0aut} in
order to define a 0-automatic queue of type $GI/GI/(X,\Sigma)$, resp.
$G/G/(X,\Sigma)$.  Roughly, the description would go as follows. The
buffering mechanism is kept unchanged; the sequence of inter-arrival
times and classes of customers is i.i.d. (resp. stationary and ergodic);
the sequence of service times at the server is i.i.d. (resp. stationary
and ergodic) and independent of the arrivals.

\subsection{Comparison with other models in the literature}

Under stability condition, we will see that a 0-automatic queue has
the ``Poisson output'' property. Also, a 0-automatic queue is
``quasi-reversible'', at least in the sense of Chao, Miyazawa, and
Pinedo~\cite[Definition 3.4]{CMPi}.  
There exist many examples of queues with such properties, see for
instance Kelly~\cite{kell79} or \cite{CMPi}. 
However, 0-automatic queues are
quite different from the existing models. 

Let us detail the comparison with the models in \cite{CMPi},
see also \cite{ChMi00}. Their model is a wide generalization of
Gelenbe's G-queue with signals, batch arrivals, and batch departures. 
In a sense, 0-automatic queues can also be viewed as a wide
generalization of G-queues. Other common features between the models 
include: non-linear traffic equations, an output rate different from
the input rate, and subtle boundary conditions to get a product form. 
Despite these similarities, the models are quite orthogonal. 
One big novelty of 0-automatic queues is the possibility for
two customers to merge and create a customer with a new
type. The algebraic foundation of 0-automatic queues is another
originality. 

\medskip

It is also worth comparing the 0-automatic queue with another model for
queues introduced by Yeung and Sengupta~\cite{YeSe}, see also
He~\cite{he} (the {\em YS model} in the following). 

\begin{figure}[ht]
\[ \epsfxsize=320pt \epsfbox{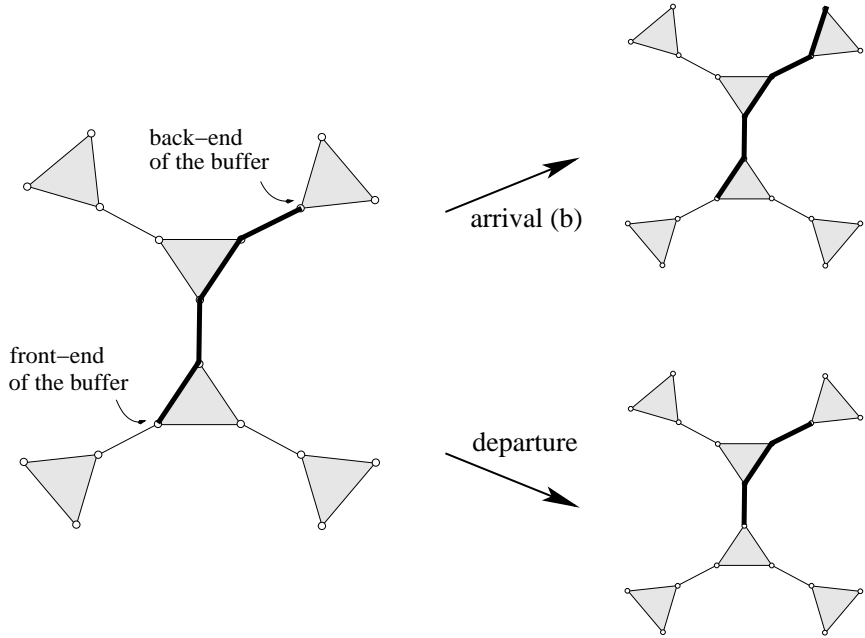} \]
\caption{Effect of an arrival and a departure on the content of the
  buffer in a 0-automatic queue built on the group $\Z/2\Z\star \Z/3\Z$.}
\label{fi-0autmeca}
\end{figure}

A common feature is the structure of the state space~: a tree for the
YS model (or the cartesian product of a tree and a finite set), and a
more general tree-like graph for the 0-automatic queue. In particular,
both models correspond to multiclass queues, and the buffer content is
coded by a word over the alphabet of classes. Second common
feature, the effect of a new arrival is either to add, to modify
the class of, or to remove, a customer at the back-end of the buffer (in
the YS model, the removal/modification may affect several customers at
the back-end of the buffer). Now, and this is the first central
difference, departures occur at the front-end of the buffer in the
0-automatic queue, and at the back-end in the YS model. Therefore, the
former is a FIFO queue while the latter is a LIFO queue. We have
illustrated the FIFO mecanism of the 0-automatic queue in Figure
\ref{fi-0autmeca}. 
The second important difference concerns the type of results which are
proved. In a stable 0-automatic queue, the buffer content has a
``product form'' stationary distribution, see Theorem \ref{th-main}. In the YS
model, it has only a  ``matrix product form'', see \cite[Section 2]{YeSe} and getting the
stronger ``product form'' requires severe additional assumptions, see 
\cite[Section 6]{YeSe}. 
To conclude the comparison, here again, the original flavor of the
0-automatic queue comes from the underlying group or monoid structure. It is
this algebraic foundation which can be accounted for the ability 
to get the strong product form results. 

\section{Stability Condition for a Zero-Automatic Queue}
\label{se-stab}

Throughout Sections \ref{se-stab} and \ref{se-main}, the model is
as follows. Let $(X,\Sigma,\nu)$ be a plain triple. Fix $\lambda$
and $\mu$ in $\R_+^*$ and $r$ in $\cb$. Consider the 0-automatic
queue $(X,\Sigma,\nu,r,\lambda,\mu)$.

\medskip

Let $M=(M_t)_t$ be the queue-content process, and $Q$ the
infinitesimal generator. Next Lemma is a direct consequence of the
strong connectivity of the graph $(\Sigma,\rightarrow)$ defined in
\eref{eq-graph}. 

\begin{lemma}
The process $M$ is irreducible.
\end{lemma}

The aim of this Section is to prove Proposition
\ref{pr-stability} which characterizes the stability region of the 0-automatic
queue.

\begin{proposition}\label{pr-stability}
Let
$\widehat{\gamma}$ be the drift of the random walk $(X,\nu)$.
We have:
\begin{eqnarray*}
 \bigl[ \lambda \widehat{\gamma} <\mu \bigr] & \iff & M \text{
   ergodic } \\
\bigl[ \lambda \widehat{\gamma} =\mu \bigr] & \iff & M \text{
   null recurrent } \\
\bigl[ \lambda \widehat{\gamma} > \mu \bigr] & \iff & M \text{
   transient} \:.
\end{eqnarray*}
\end{proposition}

Consider an excursion of $M$ from the instant $t=0$ at which it is
assumed to leave state $1_{\Sigma^*}$, to the instant $R_M$ which corresponds to
the first return to state $1_{\Sigma^*}$. Recall that $M$ is transient iff
$P\{R_M=\infty\}>0$, and ergodic iff $E[R_M]<\infty$.

It is convenient to
use the following representation for $M$.
Let $A=(A_0=0,A_1,A_2,\dots)$ where $(A_1,A_2,\dots )$ are the time
points of a time-stationary 
Poisson process of rate $\lambda$ on $\R_+$. Let $N_A =
(N_A(t))_t$ be the corresponding counting process. Let $N_D=
(N_D(t))_t$ be the counting process of a time-stationary Poisson
process of rate $\mu$ on $\R_+$.
Let $\widetilde{X}= (\widetilde{X}_n )_n$ be a realization of the
random walk $(X,\nu)$ viewed as evolving on $L(X,\Sigma)$, see
\eref{7}. Assume that $A,\widetilde{X},$ and $N_D$ are mutually
independent. Let $(X_t)_t$ be the continuous-time jump Markov
process on the state space $L(X,\Sigma)$ defined by:
\[
X_t = \widetilde{X}_{n+1} \text{ on } [A_n,A_{n+1}) \:.
\]

For all $t$ in the interval $[0,R_M)$, we have:
\begin{equation}\label{eq-repr}
|M_t|_{\Sigma} = |X_t|_{\Sigma} - N_D(t) \:.
\end{equation}

Here $X_t$ is the queue-content at time $t$ if no service has
been completed. Observe that the first letter of $X_t$
corresponds to the front-end of the buffer (the right-end in Figure
\ref{fi-0aut}), and the last letter to the back-end (the left-end in Figure
\ref{fi-0aut}).

\medskip

The counting process of a Poisson process satisfies a Strong Law
of Large Numbers. We get, a.s.,
\begin{equation}\label{eq-slln}
\lim_{t\rightarrow \infty} \frac{N_A(t)}{t} = \lambda, \qquad
\lim_{t\rightarrow \infty} \frac{N_D(t)}{t} = \mu \:.
\end{equation}

We also have, a.s.,
\begin{equation*}
\lim_{n\rightarrow \infty} \frac{|\widetilde{X}_n |_{\Sigma}}{n} =
\widehat{\gamma} \:,
\end{equation*}
where $\widehat{\gamma}$ is the drift of the random walk
$(X,\nu)$. So we have, a.s.,
\begin{equation}
\lim_{n\rightarrow \infty}  \frac{|X_t|_{\Sigma}}{t} =
\lim_{t\rightarrow \infty} \frac{|X_t|_{\Sigma}}{N_A(t)} \times
\frac{N_A(t)}{t}= \lambda\widehat{\gamma} \:.
\end{equation}


We can now prove the following.

\begin{lemma}\label{le-step0}
If $\lambda\widehat{\gamma} < \mu$ then $M$ is recurrent. If $\lambda
\widehat{\gamma} > \mu$ then $M$ is transient.
\end{lemma}

\begin{proof}
We show the first statement by contraposition. If $M$ is transient
then $P\{R_M=\infty\}>0$. Using \eref{eq-repr} and \eref{eq-slln},
we obtain that a.s. on the event $\{R_M=\infty\}$, we have:
\[
\lim_t \frac{|M_t|_{\Sigma}}{t} = \lambda \widehat{\gamma} - \mu \:.
\]
To avoid a contradiction, we must have  $\lambda \widehat{\gamma} -
\mu \geq 0$.

Now assume that $\lambda \widehat{\gamma} > \mu$. Using the
independence of $(X_t)_t,$ and $N_D$, and the regenerative properties of $R_M$, it is
easily shown that
$P\{ \forall t, \ |M_t|_{\Sigma} > 0\} >0$. In particular
$P\{R_M=\infty\}>0$ and $M$ is transient.
\end{proof}

To get the stronger results in Proposition \ref{pr-stability}, the
idea is to approximate the 0-automatic queue by a simple queue with a
Markov additive arrival process, and then to use standard results
from queueing theory.

\medskip

Since $(X_t)_t$ is transient (Proposition \ref{pr-transient}),
there exists an a.s. finite $T_0$ such that $\forall t \geq T_0,
|X_t|_{\Sigma}\geq 1$. For notational simplicity, assume that
$T_0=0$. Define the random variables: $\forall k \geq 1$,
\begin{equation}
T_k = \inf \bigl\{ t \mid \forall s\geq t, \ |X_s|_{\Sigma} \geq
k+1 \bigr\}, \qquad \tau_k = T_{k+1}-T_k \:.
\end{equation}
The r.v.'s $T_k$ are a.s. finite because $(X_t)_t$ is transient, and we
have:
$(T_0=0)< T_1<T_2 < \dots$ a.s.

By transience, $\lim_t X_t$ is a (random) infinite word on the
alphabet $\Sigma$, let us write it as $\lim_t X_t
=U_0U_1U_2\cdots$.
By definition, see Section
\ref{se-prel}, the law of $U_1U_2\cdots$ is the harmonic measure
$\nu^{\infty}$ of the random walk $(X,\nu)$. Observe that:
\[
X_{T_k^-} = U_0\cdots U_{k-1}, \qquad \forall t \geq T_k, \ X_t =
U_0\cdots U_{k-1} Y_t, \ Y_t \in L(X,\Sigma) \setminus
\{1_{\Sigma^*}\} \:.
\]
According to Theorem
\ref{th-rw}, we have: $\forall u_0\cdots u_{k-1} \in L(X,\Sigma)$,
\begin{eqnarray}
P\{U_0\cdots U_{k-1} =u_0\cdots u_{k-1}\} &  = &
\frac{\widehat{r}(u_0)}{\widehat{r}(\rig(u_0))}\cdots
\frac{\widehat{r}(u_{k-2})}{\widehat{r}(\rig(u_{k-2}))}
\widehat{r}(u_{k-1}) \nonumber \\
&  = & \widehat{r}(u_0) \frac{\widehat{r}(u_1)}{\widehat{r}(\rig(u_0))} \cdots
\frac{\widehat{r}(u_{k-1})}{\widehat{r}(\rig(u_{k-2}))} \label{eq-lawU} \:.
\end{eqnarray}

It follows that $(U_k)_k$ is a Markov chain with initial distribution
$\widehat{r}$ and transition matrix $P$ given by:
\begin{equation*}\label{eq-transi}
\forall a,b \in \Sigma, \quad P_{a,b}=\begin{cases}
\widehat{r}(b)/\widehat{r}(\rig(a)) &
                      \text{if } b\in \rig(a) \\
                      0    & \text{otherwise }
\end{cases}\:.
\end{equation*}
The matrix $P$ is irreducible as a direct consequence of the
strong connectivity of the graph of successors $(\Sigma,\rightarrow)$, see
\eref{eq-graph}. 
Let $\pi$ be the stationary
distribution of $P$ characterized by $\pi P=\pi$. In general, the
Markov chain $(U_k)_k$ is not stationary, i.e. $\widehat{r}$ is
different from $\pi$. (See \cite[Proposition 3.6]{MaMa} for a
sufficient condition on $(X,\nu)$ ensuring that
$\widehat{r}=\pi$.)

\medskip

Consider now the sequence $(U,\tau)=((U_k,\tau_k))_k$. A consequence of the above
is that $(U,\tau)$ is a Markov chain with transition function
depending only on the first coordinate. According to the classical
terminology, the sequence $(T_k)_k$ is a {\em Markov additive
  process (MAP)}.

\medskip

Consider the simple queue of type $\text{MAP}/M/1$ with arrival process
$(T_k)_k$, and a service process driven by $N$.
Let $(\sigma_k)_k$ be the corresponding sequence of service times.
We deduce from \eref{eq-slln}, that a.s. and in $L^1$:
\[
\lim_{n\rightarrow \infty} \  \frac{1}{n}\sum_{i=1}^n \bigl( \sigma_i -
\tau_i\bigr)  \ = \ \frac{1}{\mu} -
\frac{1}{\lambda \widehat{\gamma}} \:.
\]
Let $Z=(Z_t)_{t}$ be the
queue-length process of this queue. Let $R_Z$ be the
first instant of return to 0 for the process $Z$.
Applying standard results for $\text{MAP}/GI/1$ queues, see for
instance \cite[Prop. 4.2, Chapter X]{asmu87}, we get:
\begin{eqnarray}
\bigl[ \lambda \widehat{\gamma} < \mu \bigr] & \implies & \bigl[ E[R_Z]<
  \infty \bigr]  \label{eq-impli1} \\
\bigl[ \lambda \widehat{\gamma} = \mu \bigr] & \implies & \bigl[ R_Z
  \text{ a.s. finite}, \ E[R_Z]=
  \infty \bigr] \:.\label{eq-impli2}
\end{eqnarray}

Concentrating on the mechanism of the 0-automatic queue, it is not difficult
to see that:
\begin{equation}\label{eq-ZM}
\Bigl[ T_k \leq R_Z < T_{k+1} \Bigr] \implies \Bigl[ T_k \leq
 R_Z \leq
  R_M < T_{k+1} \Bigr]\:.
\end{equation}
Hence, the queue $\text{MAP}/M/1$ is a good approximation of the 0-automatic
queue. In particular, the two implications in
\eref{eq-impli1}-\eref{eq-impli2}
also hold for $R_M$. In view of Lemma \ref{le-step0}, this completes the proof of Proposition
\ref{pr-stability}.

{\bf Remark.}
The above proof does not rely in an essential way on the Markovian
assumption. For instance, modulo some care,
an analog
of Proposition \ref{pr-stability} can clearly be written for a
0-automatic queue of type $GI/GI/(X,\Sigma)$.

\section{Stationary Distribution of a Stable Queue}\label{se-main}

\subsection{The Twisted Traffic Equations}

The Traffic Equations, see Definition \ref{de-TE}, play a central role in studying
the random walk.
We now introduce equations which play a similar role
for the queue.

\begin{definition}[Twisted Traffic Equations]\label{de-TTE}
The {\em Twisted Traffic
  Equations $\text{TTE}$} associated with $(X,\Sigma,\nu,\lambda,\mu)$
are the
equations of the variables
$(\eta,x), \ \eta \in \R_+^*, \ x=(x(a))_{a\in \Sigma} \in \R_+^{\Sigma},$
defined by:
\begin{eqnarray}\label{eq-TTE}
\eta (\lambda+\mu) x(a) & = & \eta^2 \mu x(a) + \lambda \nu(a)
x(\rig(a) ) + \eta\lambda \sum_{b \ast d=a} \nu(b)x(d)
\\
& & \hspace*{2.5cm} + \eta^2 \lambda \sum_{\bdl{d\in
\lef(a)}{b\ast d=1_X}} \nu(b) \frac{x(d)}{x(\rig(d))} x(a)
\:.\nonumber
\end{eqnarray}
\end{definition}

To get a hint of the future role of  the $\text{TTE}$,
let us examine the case $(X,\Sigma)=(\{a\}^*,\{a\})$ considered at the
end of Section \ref{se-0autq}. Recall that there is only one possible
variant for the queue $M/M/(\{a\}^*,\{a\})$ which is equivalent to
the simple $M/M/1$ queue.
By simplifying \eref{eq-TTE},
we get:
\begin{equation}\label{19}
    \rho(\lambda+\mu)=\rho^2\mu+\lambda  \:.
    \end{equation}
Compare this with the global balance equations of the $M/M/1$ queue:
\begin{equation}\label{20}
    \pi(n)(\lambda+\mu)=\pi(n-1)\lambda+\pi(n+1)\mu \:.
    \end{equation}
By substituting $\pi(n)=\pi(0)\rho^n$ in (\ref{20}), we recognize
(\ref{19}).

\medskip

According to Proposition \ref{pr-solTE}, there is a unique admissible
solution to the Traffic Equations, that we denote by $\widehat{r}=
(\widehat{r}(a))_{a\in \Sigma}$. We denote by $\widehat{\gamma}$ the
drift of the random walk $(X,\nu)$.

Consider $x\in \closcb$. Define
\begin{equation}\label{eq-ABC}
A(x) = \sum_{a\in \Sigma} \nu(a) \sum_{b\in \rig(a)} x(b), \quad
B(x) = \sum_{a\ast b \in \Sigma} \nu(a)x(b), \quad C(x) =
\sum_{a\ast
  b =1_X} \nu(a)x(b) \:.
\end{equation}
One easily checks that:
\begin{equation}\label{eq-crux}
A(x)+B(x)+C(x)=1\:.
\end{equation}
This point was the crux of
the argument in proving Proposition \ref{pr-solTE} in
\cite{mair04}. Observe that a simple rewriting of \eref{eq-drift} gives:
\begin{equation}\label{eq-gammaAC}
\widehat{\gamma} = A(\widehat{r}) -C(\widehat{r}) \:.
\end{equation}

\medskip

Let us investigate some properties of the solutions to the
Twisted Traffic Equations.

\medskip

First, if $(\rho,r)$ is a solution to the TTE with $r\in \closcb$,
then $r$ belongs to $\cB$. This follows directly from the shape of
the TTE and from the strong connectivity of the graph
$(\Sigma,\rightarrow)$, see \eref{eq-graph}. 

Second, if we set $\eta=1$ in the Twisted Traffic Equations \eref{eq-TTE}, and perform the obvious
simplifications, we obtain the Traffic Equations \eref{eq-TE}.
It implies that $(1,\widehat{r})$ is a solution to the $\text{TTE}$ for
all $\lambda$ and $\mu$.

\begin{lemma}\label{le-tte1}
Let $(\rho,r), \ \rho \in \R_+^*, r\in \cb,$ be a solution to the
$\text{TTE}$. We have either $(\rho,r)=(1,\widehat{r})$, or
\begin{equation}\label{eq-etax}
\rho=\frac{\lambda\sum_{a\in \Sigma} \nu(a)r(\rig(a))}{\mu +
\lambda \sum_{a\ast b =1_X} \nu(a)r(b)} = \frac{\lambda A(r)}{\mu
+
  \lambda C(r)}\:.
\end{equation}
\end{lemma}

\begin{proof}
By summing all the Equations of \eref{eq-TTE}, we get:

\begin{eqnarray*}
\rho(\lambda+\mu) & = & \rho^2\mu + \lambda \sum_{a\in
\Sigma}\nu(a) r(\rig(a)) + \rho \lambda \sum_{
a*b \in \Sigma} \nu(a)r(b) \\
&& \hspace*{2cm} + \rho^2\lambda \sum_{\bdl{d\in
    \lef(a)}{b\ast d=1_X}} \nu(b)\frac{r(d)}{r(\rig(d))} r(a)
\\
& = & \rho^2\mu + \lambda A(r) + \rho\lambda B(r) + \rho^2\lambda
C(r) \:.
\end{eqnarray*}
Replacing $B(r)$ by $1-A(r)-C(r)$ in the above, we get:
\begin{equation}\label{23}
(\rho - 1)\bigl[ (\mu+\lambda C(r))\rho - \lambda A(r)\bigr] = 0 \:.
\end{equation}

If $\rho=1$, we have seen that the TTE reduces to the TE, which implies by
Proposition \ref{pr-solTE} that $(\rho,r)=(1,\widehat{r})$. Otherwise, we must have
$(\mu+\lambda C(r))\rho - \lambda A(r)=0$. This completes the
proof.
\end{proof}

The relevant solutions to the TTE will turn out to be the ones
satisfying \eref{eq-etax}. This leads us to the next
Definition.

\begin{definition}\label{de-admissible}
A solution $(\rho,r)$ to the TTE is called an {\em admissible solution} if
$\rho\in \R_+^*, \ r\in \cb,$ and if \eref{eq-etax} is satisfied.
\end{definition}

\begin{lemma}\label{le-tte2}
If $\lambda\widehat{\gamma}=\mu$, then $(1,\widehat{r})$
is an admissible solution to the TTE.
If $(1,r)$ is an admissible solution to the $\text{TTE}$, then
$r=\widehat{r}$ and $\lambda \widehat{\gamma} = \mu$.
\end{lemma}

\begin{proof}
Assume that $\lambda\widehat{\gamma}=\mu$.
We know that $(1,\widehat{r})$ is a solution to the TTE.
We need to check that it is admissible. By definition, it is
admissible if:
\[
1= \frac{\lambda A(\widehat{r})}{\mu + \lambda
C(\widehat{r})} \ \iff \ \lambda \bigl (A(\widehat{r})
  -C(\widehat{r})\bigr)=\mu \:.
\]
We conclude by recalling that: $\widehat{\gamma}= A(\widehat{r})
  -C(\widehat{r})$, see \eref{eq-gammaAC}.

Assume now that $(1,r)$ is an admissible solution to the $\text{TTE}$.
Since $\rho=1$, the TTE reduce to
the TE implying that  $r=\widehat{r}$. Now replacing $(\rho,r)$ by
$(1, \widehat{r})$ in \eref{eq-etax}, we get:
$\lambda\widehat{\gamma}=\mu$.
\end{proof}

More generally, admissible solutions always exist:

\begin{lemma}\label{le-goodsol}
There exists an admissible solution to the
$\text{TTE}$.
\end{lemma}

\begin{proof}
Consider the Equations \eref{eq-TTE} and replace $\eta$ by
$\lambda A(x)/(\mu+\lambda C(x))$. The resulting equations in $x$ can be
viewed as a fixed point equation of the type $\Psi(x)=x$. The
corresponding application $\Psi: (\R_+^*)^{\Sigma}\longrightarrow
(\R_+^*)^{\Sigma}$ has the following form. For $a\in \Sigma$
and for $x\in (\R_+^*)^{\Sigma}$,
\begin{eqnarray}\label{eq-phi}
\Psi(x)(a) & = & \frac{1}{\lambda + \mu} \ \Bigl[ \ \frac{ \lambda
A(x) }{\mu+\lambda
  C(x)} \bigl[ \mu x(a) + \lambda \sum_{\bdl{d\in \lef(a)}{b\ast d =1_X}} \nu(b)
  \frac{x(d)}{x(\rig(d))} x(a) \bigr] \Bigr. \nonumber \\
&& \hspace*{2cm} \Bigl. + \lambda \sum_{b\ast d=a}
  \nu(b)x(d) + \frac{\mu+\lambda C(x)}{ A(x) }
  \nu(a)x(\rig(a)) \ \Bigr]\:.
\end{eqnarray}

Consider $x\in \cb$. By summing the Equations in
\eref{eq-phi} and using \eref{eq-crux}, we get:
\[
\sum_{a\in \Sigma} \Psi(x)(a) = \frac{1}{\lambda + \mu} \ \bigl[
\lambda A(x) + \lambda B(x) + \mu+ \lambda C(x) \bigr] = 1\:.
\]
We have proved that
$\Psi(\cb)\subset
\cb$.
The end of the proof follows very closely the proof of Theorem 4.5 in
\cite{mair04}. For the sake of completeness, we recall the argument.

To use a Fixed Point Theorem, we need to define
$\Psi$ on a compact and convex set.
The set $\closcb$, which is the closure of $\cb$, is a compact
and convex subset of $\R^{\Sigma}$. But the
map $\Psi$ cannot in general be extended continuously on
$\closcb$. More precisely, $\Psi(x), x \in \closcb \moins \cb,$ can be
defined unambiguously iff $x(\rig(u)) \neq 0$ for all $u$.

For $x\in \closcb\moins \cb$, let $\Psi(x)\subset \closcb$ be the
set of possible limits of $\Psi(x_n), x_n\in \cb, x_n\rightarrow
x$. We have extended $\Psi$ to a correspondence $\Psi: \closcb
\corresp \closcb$. Clearly this correspondence has a closed graph
and nonempty convex values. Therefore, we are in the domain of
application of the Kakutani-Fan-Glicksberg Theorem, see
\cite[Chapter 16]{AlBo}. The correspondence has at least one fixed
point: $\exists r\in\closcb$ such that $r\in \Psi(r)$. Now using
the shape of the Equations in \eref{eq-phi} and the strong
connectivity of the graph of succesors $(\Sigma,\rightarrow)$, we obtain
that $r\in \cb$ (see \cite[Theorem 4.5]{mair04} for details).

Set $\rho= \lambda A(r)/(\mu +\lambda C(r))$. The pair $(\rho,r)$ is
an admissible
solution to the $\text{TTE}$.
\end{proof}

\subsection{The main results}\label{sse-main}

Next Lemma begins
to establish the link between the
Twisted Traffic Equations and the queue $M/M/(X,\Sigma)$.

\begin{lemma}\label{lemma2}
Let $(\rho,r)$ be an admissible solution to the
$\text{TTE}$. Consider the 0-automatic queue of type
$(X,\Sigma,\nu,r,\lambda,\mu)$. Let $Q_r$ be the infinitesimal generator
of the queue-content process.
Consider the measure $p_{\rho,r}$ on $L(X,\Sigma)$ defined by:
\begin{equation}\label{eq-px}
\forall a_n\cdots a_1\in L(X,\Sigma), \quad p_{\rho,r}(a_n\cdots
a_1) = \rho^n \frac{r(a_n)}{r(\rig(a_n))}\cdots
\frac{r(a_{2})}{r(\rig(a_{2}))}
    r(a_1) \:.
\end{equation}
We have $p_{\rho,r} Q_r=0$. Conversely, assume there exist
$\rho\in \R_+^*$ and $r\in \cb$ such that the measure $p_{\rho,r}$
defined by \eref{eq-px} satisfies $p_{\rho,r} Q_r=0$. Then
$(\rho,r)$ is an admissible solution to the TTE.
\end{lemma}

\begin{proof}
We have $p_{\rho,r} Q_r=0$ if and only if: $\forall u \in
L(X,\Sigma)$,
\begin{equation}\label{balance}
\sum_{\bdl{v\in L(X,\Sigma)}{v\neq u}} p_{\rho,r}(u) Q_r(u,v) =
\sum_{\bdl{v\in L(X,\Sigma)}{v\neq u}} p_{\rho,r}(v) Q_r(v,u) \:.
\end{equation}
Denote the left and right-hand side of the above equality by $L$
and $R$, respectively. Define
$$\Noact(a)=\{b \in \Sigma \mid b*a = a\} \ .$$
The left of (\ref{balance}) is:
\begin{equation}L=
\left\{%
\begin{array}{ll}
\sum_{a\in \Sigma}\lambda \nu(a) r(\rig(a))= \lambda A(r)\; & \text{if } \; u=1_{\Sigma^*} \\
p_{\rho,r}(u)(\lambda(1-\nu(\Noact(a_n))) + \mu) \; &
\mbox{otherwise }
\end{array}%
\right. \:.
\end{equation}

The right of (\ref{balance}) is given by, for $u=1_{\Sigma^*}$,
\begin{equation}
R = \sum_{a \in \Sigma} \rho r(a) \bigl[ \mu + \lambda \sum_{b\ast
a =1_X} \nu(b) \bigr] = \rho \bigl( \mu  + \lambda C(r) \bigr)\:,
\end{equation}
and for $u=a_n\cdots a_1, \ n\geq 1$,
\begin{eqnarray*}
R & = & \lambda \nu(a_n) p_{\rho,r}(a_{n-1}\cdots a_1) +
\sum_{\bdl {d \neq a_n}{b\ast d =a_n}}
\lambda \nu(b) p_{\rho,r}(da_{n-1}\cdots a_1) \\
&& \hspace*{0cm} + \sum_{\bdl {b\in \lef(a_n)}{a\ast b=1_X}}
\lambda \nu(a)p_{\rho,r}(ba_{n}\cdots a_1) + \sum_{b\in\rig(a_1)}
p_{\rho,r}(a_{n}\cdots a_1b)\mu \\
& = & p_{\rho,r}(a_n\cdots a_1) \Bigl[ \frac{1}{\rho}\frac
{r(\rig(a_n))}{r(a_n)}\lambda\nu(a_n) + \sum_{ \bdl {d\neq a_n}{b
\ast d=a_n}} \frac{r(d)}{r(\rig(d))}
\frac{r(\rig(a_n))}{r(a_n)} \lambda \nu(b) \Bigr. \\
&& \hspace*{3cm} \Bigl. + \sum_{\bdl{b\in \lef(a_n)}{a\ast b=1_X}}
\rho \lambda\nu(a) \frac {r(b)}{r(\rig(b))} + \sum_{b \in
\rig(a_1)} \rho \frac{r(b)}{r(\rig(a_1))} \mu \Bigr]\:.
\end{eqnarray*}
Now recall that  $b\ast d\in \Sigma \implies \rig(b*d)=\rig(d)$.
We obtain:
\begin{eqnarray*}
R & = & p_{\rho,r}(a_n\cdots a_1) \Bigl[ \frac{1}{\rho}\frac
{r(\rig(a_n))}{r(a_n)}\lambda\nu(a_n) + \sum_{b \ast d=a_n}
\frac{r(d)}{r(a_n)} \lambda
\nu(b) - \lambda\nu(\Noact(a_n)) \Bigr. \\
&& \hspace*{6cm} \Bigl. + \sum_{\bdl{b\in \lef(a_n)}{a\ast b=1_X}}
\rho \lambda\nu(a)\frac {r(b)}{r(\rig(b))} + \rho\mu \Bigr]\:.
\end{eqnarray*}

We see that for $u\neq 1_{\Sigma^*}$, the equality $L=R$ is
precisely equivalent to the fact that $(\rho,r)$ is a solution to
the $\text{TTE}$. For $u=1_{\Sigma^*}$, the equality $L=R$ is
precisely equivalent to the fact that $\rho$ and $r$ satisfy
\eref{eq-etax}.

Therefore, the equality $L=R$ is precisely equivalent to the fact
that $(\rho,r)$ is an admissible solution to the $\text{TTE}$.
This completes the proof.
\end{proof}

We now have all the
ingredients to prove the central results of the
paper.

\begin{theorem}\label{th-main}
Let $(X,\Sigma,\nu)$ be a plain triple. Fix $\lambda$ and
$\mu$ in $\R_+^*$.
Let $(\rho,r)$ be an admissible solution to the TTE.
Consider the 0-automatic queue
  $(X,\Sigma,\nu,r,\lambda,\mu)$.  Denote by $M_r=(M_r(t))_t$ the
  queue-content process and by $Q_r$ its infinitesimal generator.
We have:
\begin{equation*}
\begin{array}{ccccc}
[ \rho < 1 ] & \iff & [ \lambda\widehat{\gamma} < \mu
  ] & \iff & \bigl[ M_r \text{ ergodic} \bigr] \\
\left[ \rho = 1 \right] & \iff & [ \lambda\widehat{\gamma} = \mu
  ] & \iff & \bigl[ M_r \text{ null recurrent} \bigr] \\
\left[ \rho > 1 \right] & \iff & [ \lambda\widehat{\gamma} > \mu
  ] & \iff & \bigl[ M_r \text{ transient} \bigr] \:.
\end{array}
\end{equation*}
Assume that $\lambda\widehat{\gamma} <\mu$. The stationary
distribution $\pi_{\rho,r}$ of the process $M_r$ is given by:
$\forall a_n\cdots a_1 \in L(X,\Sigma)$,
\begin{equation}\label{eq-statdist}
\pi_{\rho,r} (a_n\cdots a_1) = (1-\rho)p_{\rho,r}(a_n\cdots a_1) =
(1-\rho)\rho^n q(a_n)\cdots q(a_{2}) r(a_1)\:,
\end{equation}
where $q(a)=r(a)/r(\rig(a))$ for all $a\in \Sigma$.
\end{theorem}

\begin{proof}
Let $(\rho,r)$ be an admissible solution to the Twisted Traffic
Equations. Let $p_{\rho,r}$ be the measure
defined in \eref{eq-px}.
We have (using that $\sum_{a\in \Sigma} r(a) =1$):
\[
\sum_{u\in L(X,\Sigma)} p_{\rho,r}(u) = \sum_{n\in \N}\sum_{\bdl{u\in
    L(X,\Sigma)}{|u|_{\Sigma}=n}} p_{\rho,r}(u) = \sum_{n\in \N} \rho^n \:.
\]
Hence, $\sum_u p_{\rho,r}(u)<\infty$ iff $\rho<1$. Now recall that
$p_{\rho,r} Q_r =0$, Lemma \ref{lemma2}.
It is standard (see for instance \cite[Chapter 8]{brem99})
that the process $M_r$ is ergodic iff $\sum_u
p_{\rho,r}(u) <\infty$.
Now, according to Proposition \ref{pr-stability},
$M_r$ is ergodic iff $\lambda\widehat{\gamma} <\mu$. By
combining the three equivalences, we get:
\begin{equation*}
 \bigl[\rho < 1\bigr]  \iff \bigl[ M_r \text{ ergodic } \bigr] \iff
 \bigl[\lambda \widehat{\gamma} < \mu\bigr] \:.
\end{equation*}
The result in \eref{eq-statdist} holds as a direct consequence of Lemma
\ref{lemma2}.

Now let us turn our attention to the null recurrent case. Assume that
$\rho=1$.
Using Lemma \ref{le-tte2} and Proposition \ref{pr-stability}, we have:
\begin{equation}\label{eq-deux}
\bigl[ \rho = 1 \bigr] \ \implies \  \bigl[ \lambda\widehat{\gamma} = \mu
  \bigr] \ \iff \ \bigl[ M_r \text{ null recurrent} ] \:.
\end{equation}
Let $(\widetilde{\rho},
\widetilde{r})$ be an admissible  solution to the TTE. Using the
argumentation in the forthcoming proof of
Theorem \ref{th-uniq}, we deduce that we must have
$\widetilde{\rho}=\rho=1$. Hence we have an equivalence on
the left of \eref{eq-deux}. This completes the proof.
\end{proof}



Assume that $\lambda\widehat{\gamma} = \mu$.
It follows immediately from Lemma \ref{le-tte2} and Theorem \ref{th-main}
that $(1,\widehat{r})$ is the unique admissible solution to the
TTE. We now prove a more interesting result in the same vein.

\begin{theorem}\label{th-uniq}
Consider the same model as in Theorem \ref{th-main}. Assume that
$\lambda \widehat{\gamma} < \mu$. Then the TTE have a unique
admissible solution. In particular, there is only one variant of
the 0-automatic queue $M/M/(X,\Sigma)$ with a product form
distribution.
\end{theorem}

\begin{proof}
Let $(\rho,r)$ and $(\tilde{\rho},\tilde{r})$
be two admissible solutions to the TTE.
According to Theorem
\ref{th-main}, we have $\rho<1$ and $\tilde{\rho}<1$.
Let $\pi$ and $\tilde{\pi}$ be the respective stationary distributions
of $M_r$ and $M_{\tilde{r}}$.

We now use a classical result on
ergodic Markov processes, cf for instance \cite[Chapter
  8, Theorem 5.1]{brem99}: the stationary distribution
is proportional to the time spent in each state in an excursion of the
process from $s$ to $s$, for some arbitrary state $s$.

Assume that the queue-content is $1_{\Sigma^*}$ at instant 0. Let $T$,
resp. $\tilde{T}$, be the first instant of jump of $M_r$,
resp. $M_{\tilde{r}}$.  Let $R$, resp. $\tilde{R}$, be the first
return instant to $1_{\Sigma^*}$.
We have, for all $u \in L(X,\Sigma)$ ($\sim$ stands for
`proportional to'),
\[
\pi(u) \sim E \bigl[ \int_0^R \un_{\{M_r(t)=u\}} dt \bigr], \qquad \tilde{\pi}(u)
\sim E \bigl[ \int_0^{\tilde{R}} \un_{\{M_{\tilde{r}}(t)=u\}} dt \bigr]\:.
\]
It follows that, for all $u \in L(X,\Sigma)$, $u\neq 1_{\Sigma^*}$,
\[
\pi(u) \sim E \bigl[ \int_T^R \un_{\{M_r(t)=u\}} dt \bigr], \qquad \tilde{\pi}(u)
\sim E \bigl[ \int_{\tilde{T}}^{\tilde{R}} \un_{\{M_{\tilde{r}}(t)=u\}} dt \bigr]\:.
\]
And, conditioning by the value of $M_r(T)$,
resp. $M_{\tilde{r}}(\tilde{T})$, we have, for all $u \in
L(X,\Sigma)$, $u\neq 1_{\Sigma^*}$,
\begin{eqnarray*}
\pi(u) & \sim & \sum_{a\in \Sigma} \nu(a)r(\rig(a)) \ E \bigl[
  \un_{\{M_r(T)=a\}} \int_T^R \un_{\{M_r(t)=u\}} dt \bigr] \\
\tilde{\pi}(u)
& \sim & \sum_{a\in \Sigma} \nu(a)\tilde{r}(\rig(a)) \ E \bigl[
  \un_{\{M_{\tilde{r}}(\tilde{T})=a\}}
  \int_{\tilde{T}}^{\tilde{R}} \un_{\{M_{\tilde{r}}(t)=u\}} dt \bigr]\:.
\end{eqnarray*}

Observe that the generators $Q_r$ and $Q_{\tilde{r}}$ differ only in
the line indexed by $1_{\Sigma^*}$. In other terms, the conditional law of $(M_r(t))_{t \in
[T,R]}$ on the event $\{M_r(T)=a\}$, is equal to the conditional law of $(M_{\tilde{r}}(t))_{t \in
[\tilde{T},\tilde{R}]}$ on the event $\{M_{\tilde{r}}(\tilde{T})=a\}$.
Therefore $\pi$ and $\tilde{\pi}$ are obtained as linear combinations
of the same measures $p_a$ defined by:
\[
p_a(u) = E \bigl[
  \un_{\{M_r(T)=a\}} \int_T^R \un_{\{M_r(t)=u\}} dt \bigr] = E \bigl[
  \un_{\{M_{\tilde{r}}(\tilde{T})=a\}}
  \int_{\tilde{T}}^{\tilde{R}} \un_{\{M_{\tilde{r}}(t)=u\}} dt \bigr]
  \:.
\]
For $n\in \N^*$, set $\pi(n)= \pi \{ u \in L(X,\Sigma) \mid
|u|_{\Sigma}=n\}$. Define $\tilde{\pi}(n)$ and $p_a(n)$ accordingly.
Using the above, we have, for all $n\in \N^*$,
\begin{equation}\label{eq-laststep}
\pi(n) \sim \sum_{a\in \Sigma} \nu(a)r(\rig(a)) \ p_a(n), \qquad
\tilde{\pi}(n) \sim \sum_{a\in \Sigma} \nu(a)\tilde{r}(\rig(a)) \
p_a(n) \:.
\end{equation}
Besides, according to \eref{eq-statdist}, we have $\pi(n)\sim
\rho^n$ and $\tilde{\pi}(n)\sim \tilde{\rho}^n$. In view of
\eref{eq-laststep}, we conclude easily that we must have
$\rho=\tilde{\rho}$.

It remains to prove that $r=\tilde{r}$.
Let us first show that:
\begin{equation}\label{eq-qr}
\bigl[ \forall a \in \Sigma, q(a)= \tilde{q}(a) \bigr] \implies \bigl[
  \forall a \in \Sigma, r(a)= \tilde{r}(a) \bigr]\:.
\end{equation}
We have: $\forall a, \ q(a) =r(a)/r(\rig(a))$. We can reinterpret this
as: $Mr=r$, with $r$ being viewed as a column vector and with $M$
being the matrix of dimension $\Sigma\times \Sigma$ defined by:
\[
\forall a,b \in \Sigma, \quad M_{a,b} = \begin{cases} q(a) &
\text{if
  } b \in \rig(a) \\
0 & \text{otherwise}
\end{cases} \:.
\]
 Consequently, the 
matrix $M$ is irreducible. 
Now invoking the Perron-Frobenius Theorem, since $r$ has all its
coordinates positive, it implies that $r$ is necessarily the
Perron eigenvector of the matrix, i.e. the unique (up to a
multiplicative constant) eigenvector associated with the spectral
radius. But we also have: $M\tilde{r}=\tilde{r}$, and $\sum_{a}
r(a) =\sum_{a}\tilde{r}(a) =1$. By uniqueness of the Perron
eigenvector, we conclude that $r=\tilde{r}$.

\medskip

Therefore, it remains to prove that: $\forall a\in \Sigma, \
q(a)=\tilde{q}(a)$.
Define $C(X,\Sigma)\subset L(X,\Sigma)$ by:
\[
C(X,\Sigma) = \{ u_1\cdots u_k\in
L(X,\Sigma) \mid u_1\in \rig(u_k) \}\:.
\]
Observe that: $[u\in C(X,\Sigma)] \implies [\forall n, \ u^n \in
  C(X,\Sigma)]$.

For $u=u_1\cdots u_k\in C(X,\Sigma)$, set $q(u)=q(u_1)\cdots
q(u_k)$. Define $\tilde{q}(u)$ analogously. Using
\eref{eq-statdist}, we have, for all $n\in \N^*$, for some
constants $C_1$, $C_2$,
\begin{equation}\label{eq-ref}
\pi(u^n) =C_1\rho^{n|u|_{\Sigma}}
q(u)^nr(\rig(u_k)),\qquad \tilde{\pi}(u^n) =C_2
\rho^{n|u|_{\Sigma}} \tilde{q}(u)^n \tilde{r}(\rig(u_k))\:.
\end{equation}
Besides, $\pi(u^n) = C_3\sum_{a\in \Sigma} \nu(a)r(\rig(a))
p_a(u^n)$ and $\tilde{\pi}(u^n) = C_4\sum_{a\in \Sigma}
\nu(a)\tilde{r}(\rig(a)) p_a(u^n)$, for some constants $C_3$,
$C_4$. Therefore, $\pi(u^n)$ and $\tilde{\pi}(u^n)$ must grow at the
same exponential speed as a function of $n$. In view of \eref{eq-ref},
we must have: 
\begin{equation}\label{eq-circ}
q(u_1)\cdots q(u_k)=\tilde{q}(u_1)\cdots \tilde{q}(u_k) \:.
\end{equation}

For the remaining step, the argument depends on the form of $X$.
Set $X= X_1\star \cdots \star X_K \star X_{K+1} \star F $, with
$X_{K+1}=\Sigma_{K+1}^*$($|\Sigma_{K+1}|\geq 0$), with
$F=\F(\Sigma_{K+2})$ ($|\Sigma_{K+2}|\geq 0$), with $X_i$ being
finite monoids for $1\leq i \leq K$, and with
$\Sigma=\Sigma_{K+2}^{-1}\sqcup_{1\leq i\leq K+2} \Sigma_i, \
\Sigma_i=(X_i\moins\{1_{X_i}\})$ for $1\leq i \leq K$.

For $a\in\Sigma_i$, $1\leq i \leq K$, $\rig(a)=\Sigma \setminus
\Sigma_i$. For $b \in \Sigma_{K+1}$, $\rig(b)=\Sigma$. For
$c\in\Sigma_{K+2}$, $\rig(c)=\Sigma\setminus\{c^{-1}\}$. 

\medskip

We first treat the case $|\Sigma_{K+1}| + |\Sigma_{K+2}| \geq 1$.
Then there exists $i \in \{K+1,K+2\}$ such that $|\Sigma_{i}|\geq
1$. Set $\text{Gen}=\Sigma_{K+1} \sqcup
\Sigma_{K+2}\sqcup\Sigma_{K+2}^{-1}$. 
For all $a\in \text{Gen}$, $a \in \rig(a)$, so we have
$q(a)^2=\tilde{q}(a)^2$. It implies that $q(a)=\tilde{q}(a)$.
Consider $b\in\Sigma\setminus \text{Gen}$ and $a\in \text{Gen}$, one
has $b\in\rig(a)$, so, according to \eref{eq-circ},
$q(b)q(a)=\tilde{q}(b)\tilde{q}(a)$. Hence, $q(b)=\tilde{q}(b)$.
So $q(c)=\tilde{q}(c)$ for all $c\in \Sigma$.

\medskip

We now consider the case $|\Sigma_{K+1}| + |\Sigma_{K+2}| =0$.
Assume that $K\geq 3$. Consider $a\in \Sigma_1, b\in \Sigma_2,
c\in \Sigma_3$. We have: $\rig(a)=\Sigma\moins \Sigma_1$,
$\rig(b)=\Sigma\moins \Sigma_2$, and $\rig(c)=\Sigma\moins
\Sigma_3$. Therefore, $abc \in C(X,\Sigma)$ and $bc \in
C(X,\Sigma)$. Using \eref{eq-circ}, we deduce that:
$q(a)q(b)q(c)=\tilde{q}(a)\tilde{q}(b)\tilde{q}(c)$ and
$q(b)q(c)=\tilde{q}(b)\tilde{q}(c)$. We conclude that
$q(a)=\tilde{q}(a)$ for all $a\in \Sigma$.

\medskip

Assume now that $K=2$. The above argument does not work anymore.
First of all, we want to prove that:
\begin{equation}\label{eq-prop}
\forall a\in \Sigma_i, \quad \frac{r(a)}{r(\Sigma_i)} =
\frac{\tilde{r}(a)}{\tilde{r}(\Sigma_i)} \:.
\end{equation}
Consider $a,b \in \Sigma_1$ and $c\in \Sigma_2$. We have:
$\rig(a)=\rig(b)=\Sigma_2$ and $\rig(c)=\Sigma_1$. Hence, $ac \in
C(X,\Sigma)$ and $bc \in C(X,\Sigma)$. Using \eref{eq-circ}, we
have: $q(a)q(c)=\tilde{q}(a)\tilde{q}(c)$ and $
q(b)q(c)=\tilde{q}(b)\tilde{q}(c)$. It implies that:
$q(a)/q(b)=\tilde{q}(a)/\tilde{q}(b)$, which is equivalent to
\eref{eq-prop}. Set
$R(a)=r(a)/r(\Sigma_i)=\tilde{r}(a)/\tilde{r}(\Sigma_i)$ if $a\in
\Sigma_i$.

Now let us sum the TTE corresponding to all the elements of
$\Sigma_1$, and let us perform the simplifications implied by
\eref{eq-prop}. For instance for $(\rho,r)$, we get:
\begin{eqnarray*}
\rho(\lambda+\mu) r(\Sigma_1) & = &  \rho^2 \mu r(\Sigma_1) +
\lambda \nu(\Sigma_1) r(\Sigma_2) + \rho\lambda \sum_{a \ast b\in
\Sigma_1}
 \nu(a)r(b) \\
&& \hspace*{4cm} + \ \rho^2 \lambda \sum_{a\in \Sigma_1}
\sum_{\bdl{b,d\in \Sigma_2}{b\ast
     d=1_{X_2}}} \nu(b)\frac{r(d)}{r(\Sigma_1)}r(a)  \\
 & = &  \rho^2 \mu r(\Sigma_1) + \lambda
\nu(\Sigma_1) r(\Sigma_2) + \rho\lambda r(\Sigma_1)\sum_{a \ast
b\in
  \Sigma_1}  \nu(a)R(b) \\
&& \hspace*{4cm} + \ \rho^2 \lambda r(\Sigma_2) \sum_{\bdl{b,d\in
\Sigma_2}{b\ast
     d=1_{X_2}}} \nu(b) R(d)
\end{eqnarray*}
Set $B_1 = \sum_{a \ast b\in
  \Sigma_1}  \nu(a)R(b)$ and $C_2= \sum_{a,b\in \Sigma_2, \ a\ast
     b=1_{X_2}} \nu(a) R(b)$.
Using that $r(\Sigma_2)=1-r(\Sigma_1)$, we get:
\begin{equation}\label{eq-ouf}
r(\Sigma_1)  =  \frac{\lambda\nu(\Sigma_1) + \rho^2\lambda
  C_2}{\rho(\lambda+\mu) -\rho^2\mu + \lambda \nu(\Sigma_1) -\rho
  \lambda B_1 +\rho^2\lambda C_2} \:.
\end{equation}
But all the terms in the right-hand side of \eref{eq-ouf} are
unchanged when we write the corresponding equation for $\tilde{r}$.
So, $r(\Sigma_1)=\tilde{r}(\Sigma_1)$. In view of \eref{eq-prop}, we
deduce that $r(a)=\tilde{r}(a)$ for all $a\in \Sigma$.
This completes the proof. 
\end{proof}

When the triple is not plain, the TTE may have several
admissible solutions. This is for instance the case for the triple
$(\F(a),\{a,a^{-1}\},\{1/2,1/2\})$ as discussed in Section
\ref{ssse-fg}. 

\remark In the case $\rho<1$, if the boundary condition is chosen
according to $r'$, where $(\rho,r')$ is not a solution to the TTE,
then the stationary distribution of $M_{r'}$ exists
(Prop. \ref{pr-stability}). But we do not know 
how to compute it exactly. 
See Remark \ref{rm-bc} for a
justification of the form of the boundary condition.


\paragraph{Poisson departure processes.} $ $ \\


The celebrated Burke Theorem states that the departure process from a
stable $M/M/1$ queue is a Poisson process of the same rate as the
arrival process.
A nice consequence of Theorem \ref{th-main} is that an analog
of Burke Theorem holds for 0-automatic
queues.

\medskip

In a 0-automatic queue, `departures' occur both at the front-end and
at the back-end of the buffer. Here we consider only the front-end
departures, i.e. the ones corresponding to service completions and not
to buffer cancellations.

Let $M=(M(t))_t$ be the queue-content process of some 0-automatic
queue $M/M/(X,\Sigma)$. A {\em departure} is
an instant of jump of $M$ corresponding to a jump of the type:
$u_n\cdots u_1 \rightarrow u_n\cdots u_2$ for $u=u_n\cdots u_1 \in
L(X,\Sigma)\setminus \{1_{\Sigma^*}\}$.
When $u=a^n, a\in \Sigma, n\geq 1$ (the case \eref{eq-main2} in Definition
\ref{de-0aut}), some special care must be taken. The jumps of type
$a^n \rightarrow a^{n-1}$ which are {\em departures} occur at rate
$\mu$.
The {\em departure process} is the point process of
departures.

\begin{theorem}\label{th-burke}
The model is the same as in Theorem \ref{th-main}. Assume that
$\lambda\widehat{\gamma} < \mu$. Let $(\rho,r)$ be an admissible solution
of the $\text{TTE}$. Consider the 0-automatic queue
$(X,\Sigma,\nu,r,\lambda,\mu)$.
The stationary departure process is a Poisson process of rate $\rho\mu$.
Furthermore, for all $t$, the queue-content at time $t$ is
independent of the departure process up to
time $t$.
\end{theorem}

\begin{proof}
The simplest proof of Burke Theorem uses reversibility and is due to
Reich, see for instance \cite{kell79} for details. Here, the argument
is similar.

Let $M_r$ be the stationary queue-content process. Its marginal
distribution at a given instant is $\pi_{\rho,r}$ given in
\eref{eq-statdist}. Let $D$ be the corresponding departure
process. By definition, the instantaneous rate of $D$ is
$c(t)=\mu$ if $M_r(t)\neq 1_{\Sigma^*}$ and $c(t)=0$ otherwise.

Now let us consider the time-reversed point process
$\widetilde{D}$. The process $\widetilde{D}$ corresponds to the
instants of ``right-increase'' of the time-reversed process
$(\widetilde{M}_r(t))_t$. Therefore, the instantaneous rate
$\tilde{c}(t)$ of $\widetilde{D}$ is as follows.
If $\widetilde{M}_r(t)=a_n\cdots a_1 \in L(X,\Sigma) \moins \{1_{\Sigma^*}\}$,
\begin{equation}\label{eq-burke}
\tilde{c}(t) = \sum_{a\in \rig(a_1)} \frac{\pi_{\rho,r}(a_n\cdots a_1
  a)}{\pi_{\rho,r}(a_n\cdots a_1)} \mu = \sum_{a\in \rig(a_1)}
  \frac{\rho r(a)}{r(\rig(a_1))} \mu =\rho \mu \:,
\end{equation}
and if $\widetilde{M}_r(t)=1_{\Sigma^*}$,
\[
\tilde{c}(t) = \sum_{a\in \Sigma} \frac{
  \pi_{\rho,r}(a)}{\pi_{\rho,r}(1_{\Sigma^*})} \mu = \sum_{a\in \Sigma} \rho
  r(a) \mu = \rho \mu \:.
\]
We conclude that $\widetilde{D}$ is a Poisson process of rate
$\rho\mu$. Since Poisson processes are preserved by time-reversal, $D$
is also a Poisson process of rate $\rho\mu$.

Also, using the Markov property of Poisson processes,
$\widetilde{M}_r(t)$ is independent of the process $\widetilde{D}$
after time $t$. Under time-reversal, this translates as: $M_r(t)$ is
independent of the departure process $D$ up to time $t$.
\end{proof}


Here are some additional comments on Theorem \ref{th-burke}.

\medskip

1- The infinitesimal generator of the time-reversed process
$\widetilde{M}_r$ is certainly not the infinitesimal generator of
a 0-automatic queue. This was already the case for the G-queue. But
this is in contrast with the situation for the M/M/1 queue. 

\medskip

2- For $a\in \Sigma$, the departure process $D_a$ of
customers of class $a$ is not a Poisson process.

\medskip

3- Equation \eref{eq-burke} corresponds to the condition defining
``quasi-reversibility'' in Chao, Miyazawa, and
Pinedo~\cite[Definition 3.4]{CMPi}. 

\medskip

4- The saturation principle of Baccelli and Foss~\cite{BaFo93b} holds
for many classical queueing systems. Here is a  rough description of
it.

Consider a queueing system with an infinite capacity
buffer.
Let $\mu_0$ be the departure rate in the {\em saturated system} in
which an infinite number of
customers are stacked in the buffer.
Now, if the actual arrival rate in the system is $\lambda<\mu_0$, then the system is
stable, and the departure rate is $\lambda$.
A dual presentation of the same principle is as follows. Let $\lambda_0$
be the growth rate of the buffer in the {\em blocked system} where the server has
been shut down.
If the actual service rate in the system is $\mu > \lambda_0$, then the
system is stable, and the departure rate is $\lambda_0$.

Zero-automatic queues do {\em not} satisfy the saturation principle.
This can be viewed on the following inequalities (to be deduced from
Theorem \ref{th-burke}):
\[
\lambda \widehat{\gamma} \leq \rho \mu < \mu\:,
\]
where: $\rho\mu$ is the actual
 departure rate in equilibrium, $\mu$ is the departure rate from the
 saturated system, and $\lambda\widehat{\gamma}$ is the
 growth rate of the buffer in the blocked system.

\subsection{Quasi-Birth-and-Death processes}

Quasi-Birth-and-Death (QBD) processes appear naturally in the
modelling of several queueing and
communication systems. As such, they have been extensively studied,
see for instance the monographs \cite{LaRa,neut}.
The results in Section \ref{sse-main} can be put in perspective by
considering the relation between 0-automatic queues and QBD
processes.

\medskip

To that purpose, we define an  ``approximated'' and quite
simplified version of the 0-automatic queue. The idea is to keep
track of the queue-content only through the number of customers
and the class of the back-end customer. Clearly a difficulty
arises: if a cancellation occurs at the back-end of the buffer,
there is no way to retrieve the class of the new back-end
customer. This missing information is compensated as follows: the
class is chosen at random according to the relevant conditional
law.

\medskip

Consider a 0-automatic queue. The notations and assumptions are the
ones of Theorem \ref{th-main}. In particular $(\rho,r)$ is an
admissible solution to the TTE. We assume that $\lambda
\widehat{\gamma}<\mu$.  Recall that $Q_r$ is the infinitesimal
generator of the queue-content on the state space
$L(X,\Sigma)$, and that $\pi_{\rho,r}$ given by \eref{eq-statdist} is its stationary
distribution.

Consider the application:
\[
\begin{array}{cccc}
f: & L(X,\Sigma) & \longrightarrow & \{0\}\cup (\N^*\times \Sigma) \\
& u_n\cdots u_1 & \longmapsto & (n,u_n) \\
& 1_{\Sigma^*} & \longmapsto & 0
\end{array}
\]
Define the infinitesimal generator $\widetilde{Q}_r$ on the state space
$\{0\}\cup (\N^*\times \Sigma)$ by:
\begin{equation}\label{eq-tildeQ}
\widetilde{Q}_r(x,y) = \sum_{u\in f^{-1}(x)}
\frac{\pi_{\rho,r}(u)}{\pi_{\rho,r}(f^{-1}(x))} \sum_{v\in f^{-1}(y)}
Q_r(u,v) \:.
\end{equation}
For instance, we obtain by using \eref{eq-statdist} and simplifying:
\[
\widetilde{Q}_r\bigl( (n,a); (n-1,b) \bigr) = \lambda  \sum_{c\ast
  a=1_X} \nu(c) \frac{r(b)}{r(\rig(a))}\un_{\{b \in \rig(a)\}} \:.
\]

Define a total order on $\{0\}\cup (\N^*\times \Sigma)$ as
follows: $0$ is the smallest element and $(n,a)\leq (m,b)$ if $n<
m$ or $(n=m, a\preceq b)$, where $\preceq$ is some total order on
$\Sigma$. A couple of lines of computation enable to check the
following. If lines and columns are ranked according to the above
order, the infinitesimal generator $\widetilde{Q}_r$ is block
tridiagonal of the form:
\begin{equation}\label{eq-QBD}
\widetilde{Q}_r=
\left(%
\begin{array}{ccccc}
  a & A(r) & 0 & 0 &  \\
  B & A_1 & A_0 & 0 & \ddots \\
  0 & A_2(r) & A_1 & A_0 & \ddots \\
  0 & 0 & A_2(r) & A_1 & \ddots \\
   & \ddots & \ddots & \ddots & \ddots \\
\end{array}%
\right) \:,
\end{equation}
where $a$ is of dimension $1\times 1$, $A(r)$ is of dimension $1\times \Sigma$, $B$ is
of dimension $\Sigma \times 1$, and $A_0,A_1,A_2(r)$ are of dimension
$\Sigma\times \Sigma$. Furthermore the entries of $B,A_0,A_1$ can be
expressed in function of $\lambda,\mu$, and $\nu$; and the entries of
$A(r),A_2(r)$ can be expressed in function of
$\lambda,\mu,\nu$, and $r$.

\medskip

According to the terminology in Neuts \cite{neut}, $\widetilde{Q}_r$
is the infinitesimal generator of a QBD
process with a complex boundary behavior. For any such ergodic
process, the shape of the stationary
distribution is known, see for instance~\cite[Chapter
  1.5]{neut}. So we assume that $\widetilde{Q}_r$ is ergodic and we
apply the general results to get the stationary
distribution $\widetilde{\pi}$:
\begin{equation}\label{eq-stqbd}
\widetilde{\pi}(0) = y, \quad \forall n\geq 1,\forall a\in \Sigma, \ \widetilde{\pi}(n,a) =
(xR^{n-1})(a) \:,
\end{equation}
where
\begin{equation}\label{eq-stqbd2}
\begin{cases}
A_0 + RA_1 + R^2 A_2(r) = 0 & \\
ay + xB =0, \ yA(r) + x\bigl( A_1 + RA_2(r) \bigr) = 0 & \\
y + x (I -R)^{-1} (1,\dots ,1)^T =1 &
\end{cases}\:.
\end{equation}
In \eref{eq-stqbd2}, $R$ is a matrix of dimension $\Sigma\times \Sigma$ and
is the minimal nonnegative solution to the first Equation. The pair
$(y,x)$, where $y$ is a scalar and $x$ is a line
vector of dimension $\Sigma$, is the unique positive solution to
the second and third Equations.

\medskip

The stationary distribution $\widetilde{\pi}$ in \eref{eq-stqbd} has a
{\em matrix product form}. 
This matrix product form is said to be a {\em product form} (this
is called {\em Level-Geometric with parameter 0} in \cite{DaQu}) if:
$\widetilde{\pi}(n,a)=\eta^{n-1}\widetilde{\pi}(1,a)= \eta^{n-1}x(a)$, for some $\eta\in
(0,1)$. Clearly, a necessary and sufficient condition for this to hold
is: $xR=\eta x$.  Assume that this last equality holds.
By multiplying the first Equation in \eref{eq-stqbd2} by $x$, and by
simplifying the Equations, we get:
\begin{equation}\label{eq-stqbd3}
\begin{cases}
x\bigl( A_0 + \eta A_1 + \eta^2 A_2(r)\bigr) = 0 & \\
ay + xB =0, \ yA(r) + x\bigl( A_1 + \eta A_2(r) \bigr) = 0 & \\
y + (1-\eta)^{-1} x (1,\dots ,1)^T =1 &
\end{cases}\:.
\end{equation}
Now let us replace $x$ by $cr, c\in \R,$ and $\eta$ by $\rho$ in
\eref{eq-stqbd3}. The first Equation yields precisely the Twisted
Traffic Equations \eref{eq-TTE} for the pair $(\rho,r)$. The other two Equations
yield exactly: $y=1-\rho$, $c=\rho(1-\rho)$.

We conclude that the stationary distribution of
$\widetilde{Q}_r$ is given by:
\begin{equation}\label{eq-stfinal}
 \widetilde{\pi}(0) = 1-\rho, \quad \forall n\geq 1, \forall a\in
 \Sigma, \ \widetilde{\pi}(n,a)=(1-\rho)\rho^n r(a) \:.
\end{equation}

This is coherent with the form of $\pi_{\rho,r}$ as given in
\eref{eq-statdist}.
So the above provides an a-posteriori and partial justification for the shape of $\pi_{\rho,r}$
It also provides another light on the central role of the TTE.

It is classical in QBD theory~\cite{LaRa,DaQu} to modify the boundary condition to
get a stationary distribution of product form. Here the situation is
more complex since not only the
boundary condition, but also the $A_2$ matrix, are modified in the
quest for the product form.

Observe also that the result in Theorem \ref{th-main} is
much deeper than the one in \eref{eq-stfinal}. In particular, there
is a-priori no way to retrieve the result on $Q_r$ from the one on the
simplified generator $\widetilde{Q}_r$.


\section{Extension and Examples}\label{se-examples}

\subsection{Zero-automatic queues built on 0-automatic pairs}\label{sse-G}

All the results in Sections \ref{se-stab} and \ref{se-main} are
derived for queues built on plain monoids not isomorphic to $\Z$ 
(Definition \ref{de-0aut3}). 
However, the 0-automatic queue built on $\Z$ is interesting in itself
(it corresponds to Gelenbe's G-queue, see Section \ref{se-0autq}) and
exhibits new phenomena. It is therefore worthwhile to determine the subset of the above
results which remain true for this queue. 

In fact, we define a more general framework, and the notion of
0-automatic triples extending the plain triples of Definition
\ref{de-0aut3}. 

\medskip

Let $(X,\ast)$ be a group or monoid with set of generators
$\Sigma$. Denote by $\pi : \Sigma^{*} \rightarrow X $ the monoid
homomorphism which associates to a word $a_{1}\cdots a_{k}$ of
$\Sigma^*$ the
element $a_{1}\ast \cdots \ast a_{k}$ of $X$. A language $L$ of
$\Sigma^{*}$ is a {\em cross-section} of $X$
if the restriction of $\pi$ to $L$ is a bijection. The inverse map $\Phi:
X \rightarrow L$ is then called the {\em normal form} map.
Define the language $L(X,\Sigma) \subset \Sigma^{*}$ as in 
\eref{eq-loca}. Define the sets: $\forall a \in \Sigma$,
\begin{equation}
\leftt(a) = \bigl\{ b \in \Sigma \mid b \ast a \notin \Sigma \cup
\{1_{X}\}\bigr\}, \quad \rightt(a) = \bigl\{ b \in \Sigma \mid a \ast b
\notin \Sigma \cup \{1_{X}\}\bigr\}\:.
\end{equation}
In the case of a plain monoid with natural generators, we have 
$\leftt=\rightt=\rig$, see \eref{eq-ref1}. 

\begin{definition}\label{de-0autgroup}
Let $(G,\ast)$ be a group with finite set of generators $\Sigma$.
We say that the pair $(G,\Sigma)$ is 0-\textrm{automatic} if
$L(G,\Sigma)$ is a cross-section of G.
\end{definition}

Such pairs were first considered by Stallings~\cite{stal} under
another name. It can be proved using the results from
Stallings that $G$ is necessarily isomorphic to a plain group. 
However the set $\Sigma$ may be larger than a natural (see Section \ref{sse-pmg})
set of generators of the plain group. 

\medskip

Now let us extend the notion to monoids. To get good properties it is
necessary to choose a more complex definition proposed in
\cite{mair04}. 

\begin{definition}\label{de-0autmonoid}
Let $(M,\ast)$ be a monoid with finite set of generators $\Sigma$.
Assume that $L(M,\Sigma)$ is
a cross-section. Let $\Phi : M \rightarrow L(M,\Sigma)$ be the
corresponding normal form map. Assume that: $\forall u \in M$ s.t. 
$\Phi(u) = u_{1}\cdots u_{k}$, $\forall a \in \Sigma$,

\begin{equation}\label{eq-monoid1}
\Phi(u\ast a) = \left \{ \begin{array}{ll}
u_{1}\cdots u_{k-1} \; & \mbox{if } \; u_{k} \ast a = 1_{M} \\
u_{1}\cdots u_{k-1}v \; & \mbox{if } \; u_{k} \ast a = v \in \Sigma \\
u_{1}\cdots u_{k}a \; & \mbox{otherwise}
\end{array}\right., \quad
\Phi(a\ast u) = \left \{ \begin{array}{ll}
u_{2}\cdots u_{k} \; & \mbox{if } \; a \ast u_{1} = 1_{M} \\
vu_{2}\cdots u_{k} \; & \mbox{if } \; a \ast u_{1} = v \in \Sigma \\
au_{1}\cdots u_{k} \; & \mbox{otherwise}
\end{array} \right.
\end{equation}

Assume furthermore that : $\forall a,b \in \Sigma $ such that $a
\ast b \in \Sigma$,
\begin{equation}\label{eq-monoid2}
\leftt(a\ast b)=\leftt(a),\qquad \rightt(a\ast b) = \rightt(b) \:.
\end{equation}
Then we say that the pair $(M,\Sigma)$ is 0-automatic.
\end{definition}

In the group case $M=G$, the conditions \eref{eq-monoid1} and
\eref{eq-monoid2} are implied by the fact that the language
$L(G,\Sigma)$ is a cross-section. 

\medskip

The pairs formed by a plain monoid and natural generators are
0-automatic. However, in contrast
with the group case, plain monoids do not exhaust the family
of monoids appearing in 0-automatic pairs. For instance, 
$(M=\pres{a,b}{ab=1},\{a,b\})$ is a 0-automatic pair, but the monoid $M$
is not isomorphic to a plain 
monoid. This example is studied in Section \ref{sse-5}. 

\begin{definition}\label{de-0aut4}
A triple $(X,\Sigma,\nu)$ is said to be {\em 0-automatic} if: (i)
$(X,\Sigma)$ is a 0-automatic pair with $X$ infinite; (ii) $\nu$ is a
probability measure whose support is included in $\Sigma$ and
generates $X$.
\end{definition}

Any plain triple, see Def. \ref{de-0aut3}, is 0-automatic. 
Observe that $(X,\nu)$ may not be transient in a 0-automatic
triple. Also the graph of succesors $(\Sigma,\rightarrow)$ defined in
\eref{eq-graph} may not be strongly connected. 

\medskip

Consider a 0-automatic triple $(X,\Sigma,\nu)$,  $\lambda, \mu \in
\R_+^*$, and  $r \in \closcb$, see \eref{eq-cb}. 
The {\em 0-automatic queue} of type $(X,\Sigma,\nu,r,\lambda,\mu)$ is
defined exactly as in Definition \ref{de-0aut}. 

The definition of the {\em Traffic Equations} gets modified as
follows~:
\begin{equation}\label{eq-TE2}
x(a)  =  \nu(a)x(\rightt(a)) + \sum_{b\ast d = a} \nu(b)x(d)  +
\sum_{\bdl{d\in \leftt(a)}{b\ast d=1_X}} \nu(b)
\frac{x(d)}{x(\rightt(d))} x(a) \:.
\end{equation}
The definition of the {\em Twisted Traffic Equations} is modified as well~:
\begin{eqnarray}\label{eq-TTE2}
\eta (\lambda+\mu) x(a) & = & \eta^2 \mu x(a) + \lambda \nu(a)
x(\rightt(a) ) + \eta\lambda \sum_{b \ast d=a} \nu(b)x(d)
\\
& & \hspace*{2.5cm} + \eta^2 \lambda \sum_{\bdl{d\in
\leftt(a)}{b\ast d=1_X}} \nu(b) \frac{x(d)}{x(\rightt(d))} x(a)
\:.\nonumber
\end{eqnarray}

We use the convention described after
\eref{eq-cb} to define a solution to the TTE belonging to $\closcb$.

The TE do not necessarily have a unique solution, in
contrast with
Prop. \ref{pr-solTE}.
The TTE as well do not necessarily have a unique solution, and a 
solution $r\in \closcb$ to the TTE does not necessarily belong to
$\cb$. This is in contrast with Theorem \ref{th-uniq}. 

An analog of Lemma \ref{le-tte1} holds: if $(\rho,r), r\in \closcb,$ is a
solution to the TTE then either $(\rho,r)=(1,r)$ and $r$ is a solution
to the TE, or
\begin{equation}\label{eq-etax2}
\rho= \frac{\lambda A(r)}{\mu + \lambda C(r)}\:.
\end{equation}

We can now state the following result. Observe in particuler that
there may be several variants (corresponding to different $r$'s) of
the 0-automatic queue with a product form. 

\begin{proposition}\label{pr-bis}
Let $(X,\Sigma,\nu)$ be a 0-automatic triple.
Let $(\rho,r), r\in \closcb,$ be a solution to the
$\text{TTE}$ satisfying \eref{eq-etax2}. Assume that: $\forall a \in
\Sigma, \ [
  r(a)>0] \implies [ r(\rightt(a))>0 ]$. Define
$\widetilde{\Sigma} = \{a\in \Sigma \mid r(a) >0\}$ and
$\widetilde{L}= L(X,\Sigma)\cap \widetilde{\Sigma}^*$.

Consider the 0-automatic queue of type
$(X,\Sigma,\nu,r,\lambda,\mu)$. Let $Q_r$ be the infinitesimal generator
of the queue-content process $M_r$.
Consider the measure $p_{\rho,r}$ on $\widetilde{L}$ defined by:
\begin{equation}\label{eq-pxbis}
\forall a_n\cdots a_1\in \widetilde{L}, \quad p_{\rho,r}(a_n\cdots a_1) =
\rho^n \frac{r(a_n)}{r(\rightt(a_n))}\cdots \frac{r(a_{2})}{r(\rightt(a_{2}))}
    r(a_1) \:.
\end{equation}
We have $p_{\rho,r} Q_r=0$. Besides, we have:
\begin{equation*}
\begin{array}{ccccc}
\bigl[ \lambda\widehat{\gamma} < \mu
  \bigr] & \implies & \bigl[ \rho < 1 \bigr] & \implies &
\bigl[ M_r \text{ ergodic on $\widetilde{L}$ } ] \\
\bigl[ \lambda\widehat{\gamma} > \mu
  \bigr] & \implies & \bigl[ \rho > 1
  \bigr] & \implies & \bigl[ M_r \text{ transient on $\widetilde{L}$ }  ]
\end{array}\:.
\end{equation*}
When $\lambda \widehat{\gamma}<\mu$, the stationary distribution of
$M_r$ is $\pi_{\rho,r} = (1-\rho) p_{\rho,r}$. The corresponding
stationary departure process is a Poisson process of rate $\rho\mu$.
\end{proposition}

The proofs of Lemma \ref{lemma2}, Theorem \ref{th-main}, and Theorem
\ref{th-burke} are easily
adapted to get Prop. \ref{pr-bis}.

\subsection{Five illustrating examples}\label{sse-5}

We study five particular 0-automatic queues to illustrate the
above results. We focus on three aspects: (a) the stability
region; (b) the value of the {\em load} $\rho$; (c) the existence
of several stationary regimes (for the queues built on 0-automatic
triples instead of plain triples). 

When the model is simple enough,
the TTE can be solved explicitly 
to get closed form formulas as for  $\Z/3\Z \star \Z/3\Z$ below.
In all cases and like any set of algebraic equations, the TTE can
be solved with any prescribed precision.

\medskip

Another goal is to convey the
idea that 0-automatic queues ought to be pertinent in several
modelling contexts, due to the flexibility in their definition.
The five examples below should be interpreted having in mind the 
different ``types'' of tasks detailed in the Introduction (classical,
positive/negative, ``one equals many'', and ``dating agency'').

\subsubsection{The free product $\Z/3\Z\star \Z/3\Z$} 

Consider the plain triple $(\Z/3\Z\star
\Z/3\Z,\Sigma=\{a,a^2,b,b^2\}, \nu)$, where $\nu(a)=\nu(b)=p$,
$\nu(a^2)=\nu(b^2)=q=1/2-p$, $p\in (0,1/2)$.

In \cite[Section 4.2]{MaMa}, the drift is computed, it is given
by:
$$\widehat{\gamma}=-\frac{1}{4}+\frac{1}{4}\sqrt{16p^2-8p+5}.$$

According to Theorem \ref{th-uniq}, in the stable case, the associated TTE have a
unique admissible solution that we denote by $(\rho,r)$. Solving
the TTE, we get that:
\begin{eqnarray*}
\widehat{r}(a)=\widehat{r}(b)=\frac{4\lambda p^2-2\lambda
p+4p\mu+\lambda}{4(4\lambda p^2-2\lambda
p+\mu+\lambda)}=\frac{-4pq\lambda+\lambda+4p\mu}{4(-4pq\lambda+\lambda+\mu)},
\quad \widehat{r}(a^2)=\widehat{r}(b^2)=
\frac{1}{2} - \widehat{r}(a),\\
\rho=2\frac{4\lambda^2 p^2-2\lambda^2 p
+\lambda\mu+\lambda^2}{4\lambda^2 p^2-2\lambda^2 p
+4\lambda\mu+\lambda^2+4\mu^2}.
\quad\quad\quad\quad\quad\quad\quad
\end{eqnarray*}

\begin{figure}[ht]
\[ \epsfxsize=200pt \epsfbox{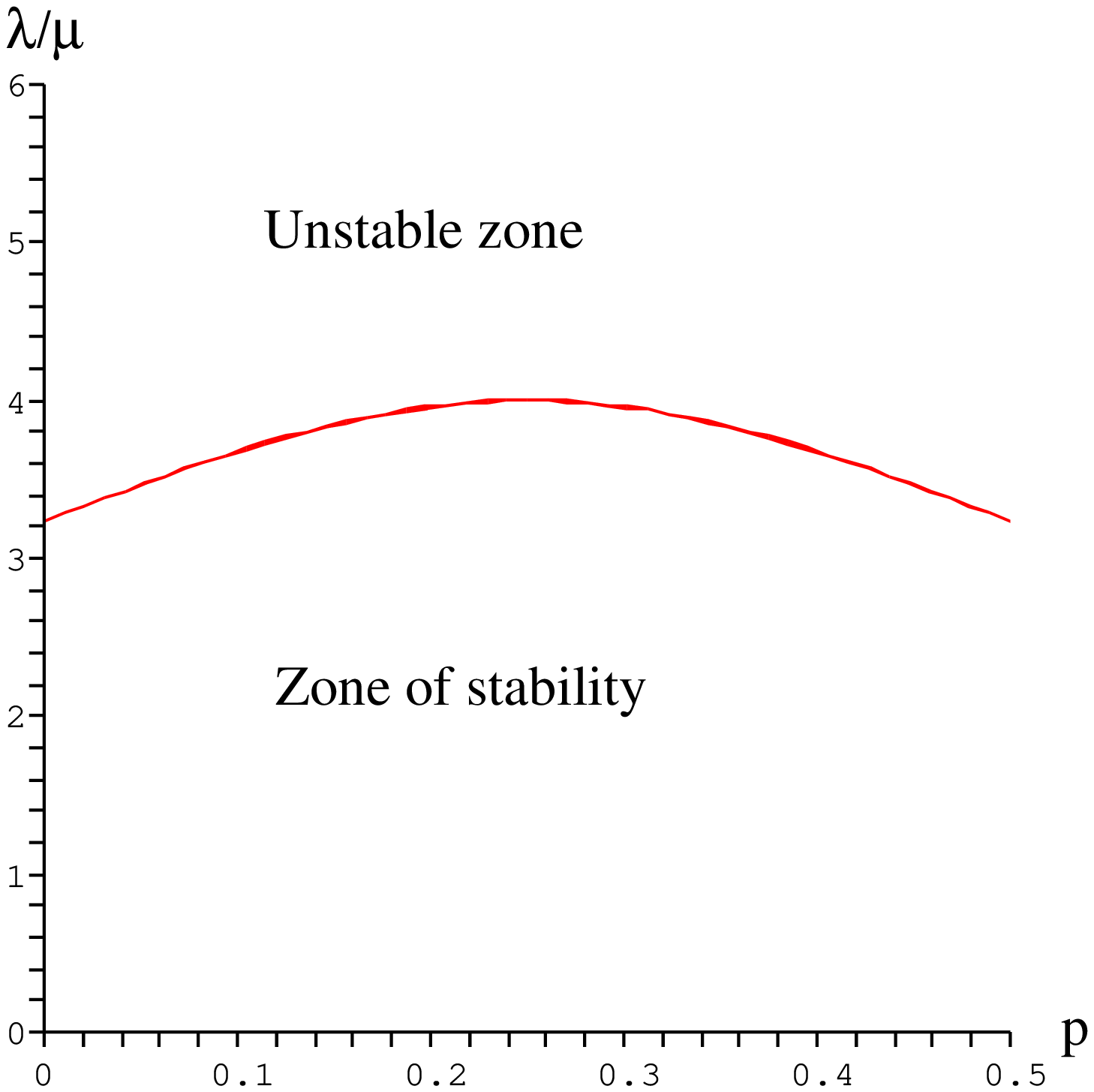} \quad \epsfxsize=210pt \epsfbox{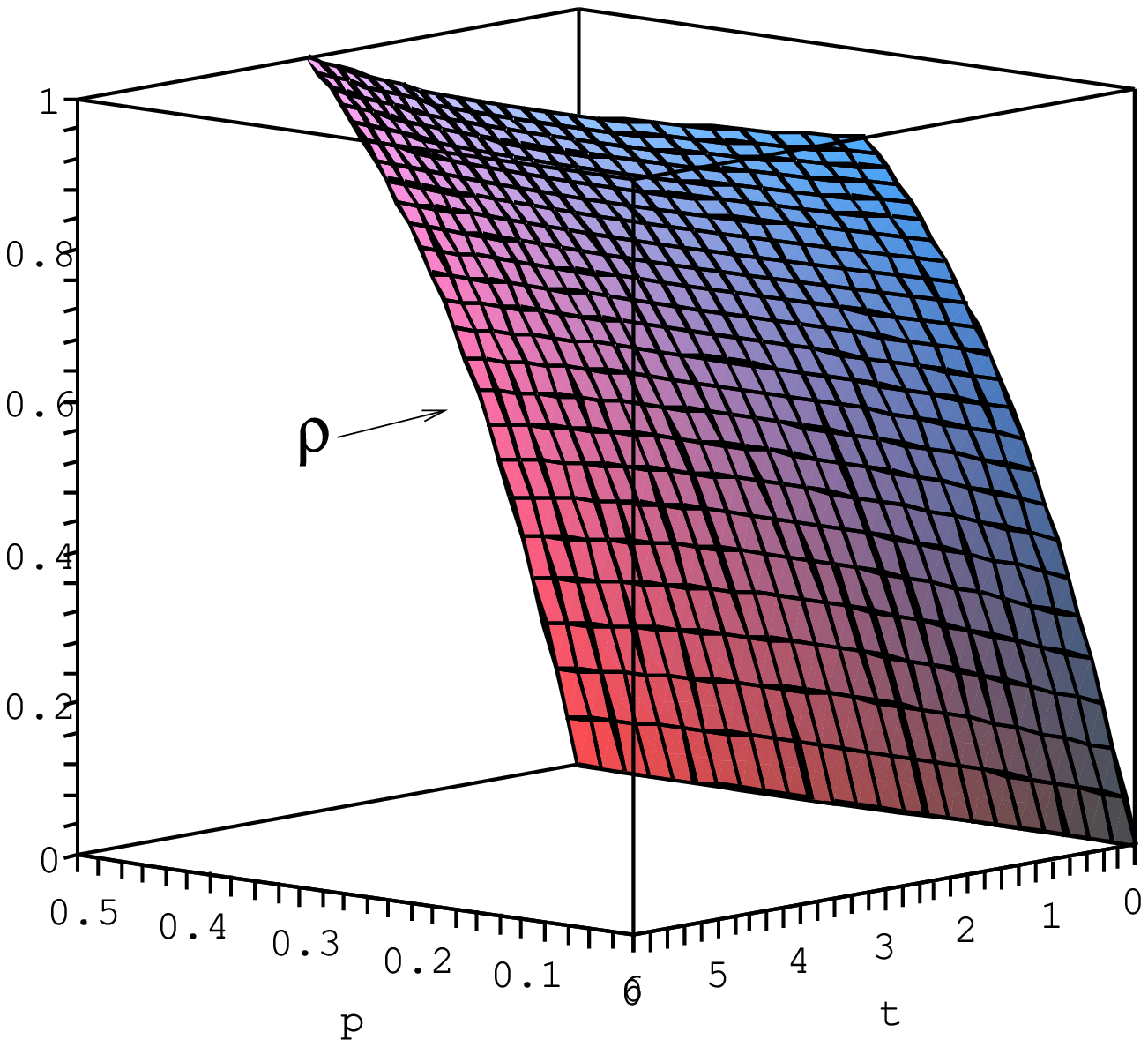} \]
\caption{$\Z/3\Z\star \Z/3\Z$: The stability region (left) and the
  load $\rho$ (right).} \label{f-z3-z3}
\end{figure}

In Figure \ref{f-z3-z3} (left), we show the stability region of
the queue. The abscissa is $p$ and the ordinate is $\lambda/\mu$.
In Figure \ref{f-z3-z3} (right), we plot the load $\rho$ as a
function of $p$ and $t=\lambda/\mu$, for $p\in (0,1/2)$ and
$\lambda/\mu \in( 0,\min(1/\widehat{\gamma},6))$. Hence, $\rho$
is always smaller or equal to 1, see Theorem \ref{th-main}.

\subsubsection{The free product $\N \star \mathbb{B}$}

\begin{figure}[ht]
\[ \epsfxsize=200pt \epsfbox{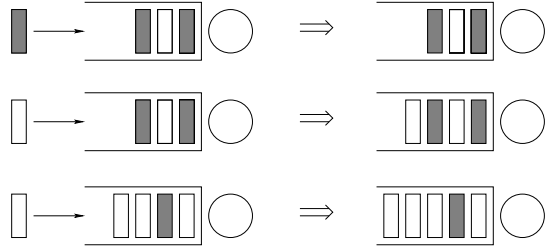} \]
\caption{The queue $M/M/(\N\star \B,\Sigma)$ with $a$ in white and
$b$ in dark gray.} \label{f-n-b}
\end{figure}

Consider the plain triple $(\{a\}^* \star
\pres{b}{b^2=b},\Sigma=\{a,b\}, \nu)$, where $\nu(a)=p$,
$\nu(b)=1-p$, $p\in (0,1)$. In Figure \ref{f-n-b}, we illustrate
the corresponding buffering mechanism.

The unique solution $\hat{r}$ of the TE is:
$\widehat{r}(c)=p $, $ \widehat{r}(b)=1-p.$ The drift of the
random walk is $\widehat{\gamma}=(2-p)p$.

According to Theorem \ref{th-uniq}, the associated TTE have a
unique admissible solution that we denote by $(\rho,r)$. Solving
the TTE, we obtain that $\rho$ is a solution of $f(Y)=0$, where:
\begin{equation*}\label{eq_rho_n-b}
f(Y)=\mu^2 Y^3+(\mu^2+\mu\lambda+\lambda\mu p)Y^2+(\lambda^2
p+\lambda\mu p)Y-\lambda^2 p^2+\lambda^2 p \:.
\end{equation*}

The relation between $r(b)$ and $\rho$ is given by: $
\rho=[r(b)(1-p)+p]\lambda/\mu$.

\begin{figure}[ht]
\[ \epsfxsize=200pt \epsfbox{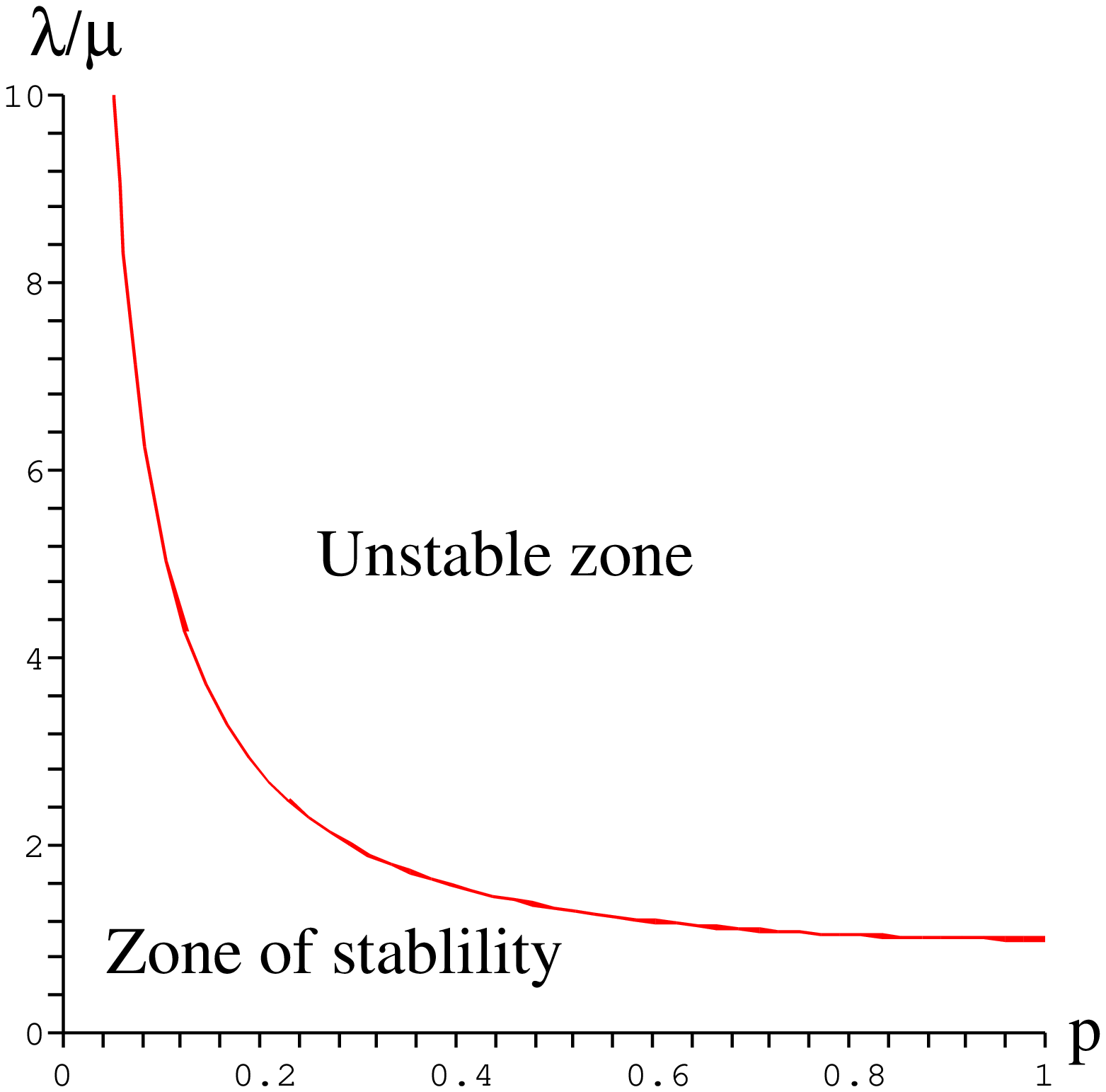} \quad \epsfxsize=220pt \epsfbox{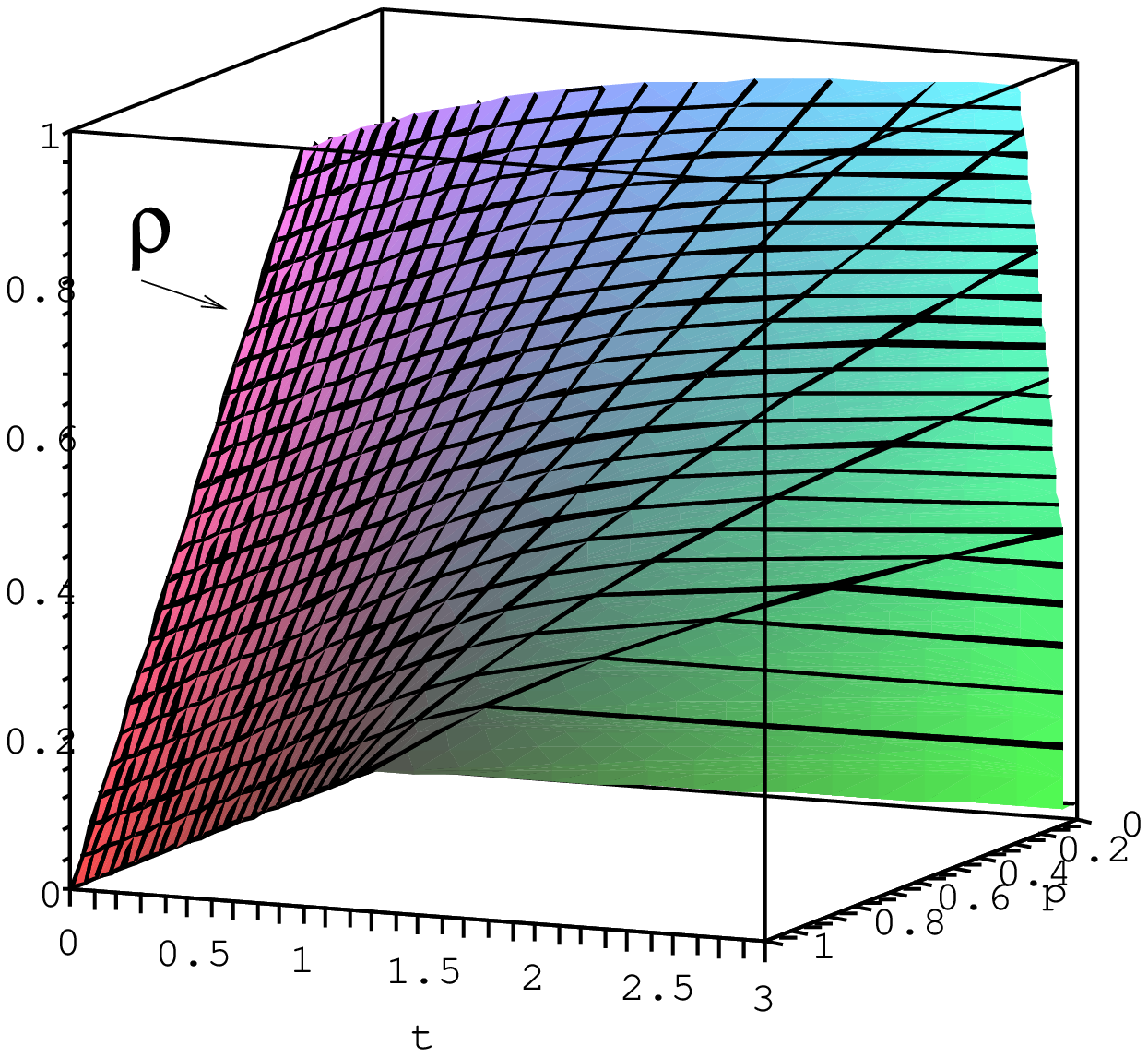} \]
\caption{$\N \star \mathbb{B}$: The stability region (left) and
the load $\rho$ (right).} \label{f-n-b-stab}
\end{figure}

In Figure \ref{f-n-b-stab} (left), we show the stability region of
the queue. The abscissa is $p$ and the ordinate is $\lambda/\mu$.
In Figure \ref{f-n-b-stab} (right), we plot the load $\rho$ as a
function of $p$ and $t=\lambda/\mu$, for $p\in (0,1)$ and
$\lambda/\mu \in( 0,\min(1/\widehat{\gamma},3))$. Hence, $\rho$ is
always smaller or equal to 1, see Theorem \ref{th-main}.

\subsubsection{The free product $\N \star \Z \star \B$}

Consider the queue associated with the plain triple $(\{a\}^*\star
\F(b)\star \pres{c}{c^2=c}, \Sigma=\{a,b,b^{-1},c\}, \nu)$ where
$\nu(a)=p, \nu(b)=\nu(b^{-1})=q/2, $ and $\nu(c) =1-p-q$ with
$p,q, p+q \in (0,1)$.

\begin{figure}[ht]
\[ \epsfxsize=220pt \epsfbox{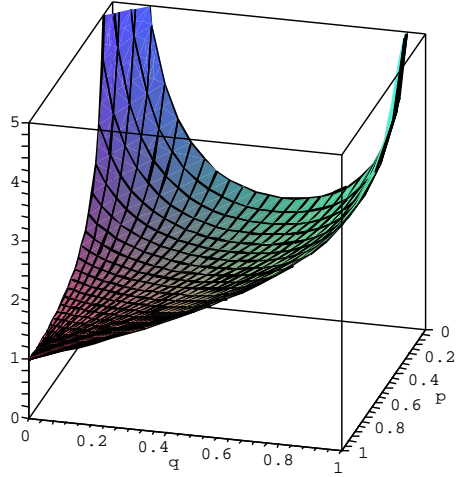} \]
\caption{Stability region of the $M/M/1/(\N\star\Z\star
  \B,\Sigma)$ queue. The axis are $p,q$, and $\lambda/\mu$. }
\label{f-n-z-b}
\end{figure}

The unique solution $\widehat{r}$ of the associated TE is
$$\widehat{r}(b)=\widehat{r}(b^{-1})=\frac{1}{2}-\frac{\sqrt{1-q^2}}{2(1+q)},
\
\widehat{r}(a)=\frac{p(1-\widehat{r}(b))}{1-\widehat{r}(b)-q\widehat{r}(b)},
\ \widehat{r}(c)=1-\widehat{r}(a)-2\widehat{r}(b).$$ Applying
Theorem \ref{th-rw}, the drift of the random walk is given by:
$\widehat{\gamma}=p+(1-p-q)(1-\widehat{r}(c))+q(1-2\widehat{r}(b))$.
From there, we obtain Figure \ref{f-n-z-b}: the stability region
is the region below the surface.

\subsubsection{The free group $\Z$ and the free product $\N \star \Z$}\label{ssse-fg}

\begin{figure}[ht]
\[ \epsfxsize=200pt \epsfbox{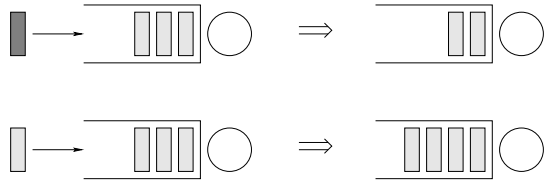} \]
\caption{The $M/M/1/(\F(a),\Sigma)$ queue with $a$ in light gray
  and $a^{-1}$ in dark gray.}
\label{fi-fg}
\end{figure}

Consider the 0-automatic queue
$(\F(a),\{a,a^{-1}\},\nu,r,\lambda,\mu)$, where $\nu$ is a
non-degenerate probability measure on $\Sigma=\{a,a^{-1}\}$. In
Figure \ref{fi-fg}, we illustrate the corresponding buffering
mechanism. Such a mechanism is similar to the one of Gelenbe's
G-queue. 

The underlying triple $(\F(a),\{a,a^{-1}\},\nu)$ is not plain
(Def. \ref{de-0aut3}), but it is 0-automatic (Def. \ref{de-0aut4}). 
Here the graph of successors $(\Sigma,\rightarrow)$, see
\eref{eq-graph}, is not connected. Also the random walk $(X,\nu)$ is
not transient but null-recurrent when $\nu(a)=\nu(a^{-1})=1/2$.

The drift of the random walk is easily computed:
\[
\widehat{\gamma} = | \nu(a) - \nu(a^{-1}) | \:.
\]

Assume first that $\nu(a)=\nu(a^{-1})$. Solving the TTE, we get
that $(\lambda/(2\mu+\lambda),r)$ is a solution for all $r \in
\closcb$. It means that the queue is stable and has a product form
distribution under any boundary condition. This interesting
behavior can be traced back to the fact that the random walk
$(X,\nu)$ is not transient.

\medskip

Assume now that $\nu(a)\neq\nu(a^{-1})$. There are 2 possible
solutions for the TTE:
\begin{equation}\label{eq-TTE-Z-1}
(\rho_1,r_1)=\bigg(
  \frac{\lambda\nu(a)}{\mu+\lambda\nu(a^{-1})};(1,0)\bigg), \qquad
(\rho_2,r_2)=\bigg(\frac{\lambda\nu(a^{-1})}{\mu+\lambda\nu(a)};(0,1)\bigg).
\end{equation}

The two solutions correspond to extremal values for $r$, it means
that in the buffer, there is only one type of customer with
probability 1: if $r_1=(1,0)$, there is only $a$ in the buffer;
if $r_2=(0,1)$, there is only $a^{-1}$ in the buffer. Here we
recover a model very close to the classical G-queue.

Set $\underline{\rho} = \min \{\rho_1,\rho_2 \}$ and $\bar{\rho}
= \max \{\rho_1,\rho_2 \}$ and define $\underline{r}$ and
$\bar{r}$ accordingly. We have:
\[
\underline{\rho} < 1, \qquad \bigl[ \bar{\rho} < 1 \bigr] \iff
\bigl[ \lambda \widehat{\gamma} < \mu \bigr] \:.
\]
The stationary distribution of the 0-automatic queue $(\F(a), \Sigma,
\nu, \underline{r},\lambda,\mu)$ is:
\begin{equation}\label{eq-1or2}
\pi_{\underline{r}} (1_{\Sigma^*})=1-\underline{\rho}\:, \quad
\pi_{\underline{r}}(x^n)=(1-\underline{\rho})\underline{\rho}^n \mbox{
  , }
\forall n\geq1,
\end{equation}
where $x=a$ if $\nu(a)< \nu(a^{-1})$, and $x=a^{-1}$ if $\nu(a) >
\nu(a^{-1})$.
When $\lambda \widehat{\gamma} < \mu$, the 0-automatic queue $(\F(a), \Sigma,
\nu, \bar{r},\lambda,\mu)$ also has a product form stationary
distribution of the form \eref{eq-1or2} with $\bar{\rho}$ instead of
$\underline{\rho}$.

\medskip

Consider now a boundary condition $r\in \cb$. In particular,
$r\neq r_1, \ r\neq r_2$, so the stationary distribution is not of
product form. However, if  $\lambda\widehat{\gamma}< \mu$, the
stationary distribution $\pi_r$ can still be determined explicitly
by solving the global balance equations. It is given by:
\begin{eqnarray}\label{eq-almost}
\pi_r(1_{\Sigma^*})& = & \Bigl( 1+r(a)\displaystyle\frac{\rho_1}
{1-\rho_1}+r(a^{-1})\displaystyle\frac{\rho_2}{1-\rho_2} \Bigr)^{-1} \nonumber\\
\pi_r(a^{n})& = & \pi_r(1_{\Sigma^*})r(a)\rho_1^n, \qquad
\pi_r(a^{-n}) \ = \ \pi_r(1_{\Sigma^*}) r(a^{-1})\rho_2^n \:,
\end{eqnarray}
where $\rho_1$ and $\rho_2$ are defined in \eref{eq-TTE-Z-1}.
The expression in \eref{eq-almost} is ``almost'' of product form. So, why do we
prefer an expression like the one in \eref{eq-1or2}~?
The point is that the departure process associated with a
stationary distribution of type \eref{eq-almost} is not Poisson, as opposed to
the one associated with \eref{eq-1or2}. And having a Poisson departure
process is crucial
to build product form networks, see \cite{DaMa06}.

\begin{figure}[ht]
\[ \epsfxsize=200pt \epsfbox{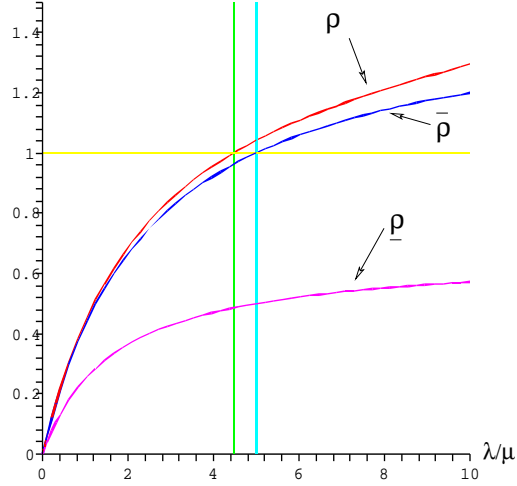} \]
\caption{$\F(a)$ and $\F(a)\star \{c\}^*$: the loads as a
  function of $\lambda/\mu$.}
\label{fi-a-petit c}
\end{figure}

To summarize, when $\lambda \widehat{\gamma} < \mu$, there are two variants of the 0-automatic
queue with a product form.
We would like to argue that one of the two
makes more ``physical'' sense.

To that purpose,
consider the plain triple $(\F(a)\star\{c\}^*,
\{a,a^{-1},c\},\nu)$ with $0< \nu(c) \ll 1$.
According to
Theorem \ref{th-uniq}, there exists a single variant of the queue with a
product form. Let $\rho$ be the corresponding load. The question is to
determine which one of the two 
solutions in \eref{eq-TTE-Z-1} is recovered when letting $\nu(c)$ go to
0.

Since the TTE are difficult to solve explicitly, we content
ourselves with numerical evidence. In Figure \ref{fi-a-petit c},
we plot $\rho$, $\bar{\rho}$, and $\underline{\rho}$ as functions
of $\lambda/\mu$, for $\nu(c)=0.01$ and $\nu(a)=p=3/5$. We see
that $\rho$ tends to the larger solution $\bar{\rho}$. The two
vertical lines correspond to the stability regions. They have an
abscissa equal to the inverse of the drift 
$\widehat{\gamma}^{-1}$ for the random walk on $\F(a)\star \{c\}^*$
and $\F(a)$ respectively.

\subsubsection{The monoid $M= \pres{a,b}{ab=1}$}

Consider the {\em bicyclic monoid} $M= \pres{a,b}{ab=1}$. Here we have
a new ``type'' 
of tasks. It is close to the positive/negative type but with no
symmetry between the positive and negative customers. 

Consider the triple $(M,\Sigma=\{a,b\},\nu)$, with
$\nu(a)=p \in (0,1)$, $\nu(b)=1-p$. It is a 0-automatic triple but not a plain
triple. In particular, the graph of successors $(\Sigma,\rightarrow)$
is not strongly connected. 

The drift of the random walk is easily
computed and given by $\widehat{\gamma}=|1-2p|$.
Solving the associated TTE, we obtain that there is one solution
if $p\leq 1/2$ and two solutions if $p> 1/2$. More precisely,
these two solutions are:
$$(\rho_1,r_1)=\bigg(\frac{\lambda(1-p)}{\mu+\lambda p};(0,1)\bigg) \mbox{ , }\forall p $$
$$ (\rho_2,r_2)=\bigg(\frac{\lambda p}{\mu+\lambda
  (1-p)};\big(\frac{2p-1}{p},
\frac{1-p}{p}\big)\bigg) \mbox{ , if } p > 1/2. $$

We have:
\begin{eqnarray*}
\text{if } p\leq  1/2, & & \bigl[ \rho_1 <1 \bigr] \iff \bigl[ \lambda
    \widehat{\gamma} < \mu \bigl] \\
\text{if } p> 1/2, &  & \rho_1 <1, \qquad  \bigl[ \rho_2 <1 \bigr] \iff \bigl[ \lambda
  \widehat{\gamma} < \mu \bigl] \:.
\end{eqnarray*}

\begin{figure}[ht]
\[ \epsfxsize=190pt \epsfbox{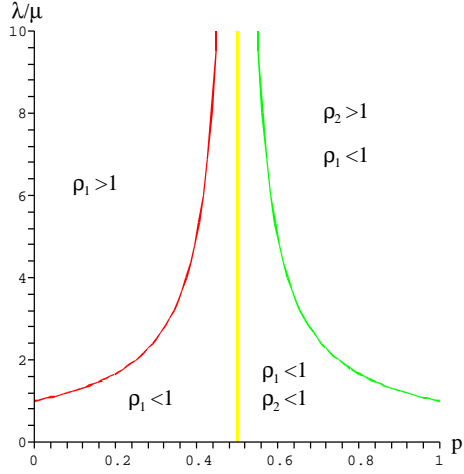} \]
\caption{$\F(a)$: The solutions to the TTE.}
\label{fi-stable_c}
\end{figure}

In Figure \ref{fi-stable_c}, we show the solutions to the TTE as a function of
$p$ and $\lambda/\mu$.

To discriminate between $(\rho_1,r_1)$ and $(\rho_2,r_2)$, we proceed
as for $\F(a)$.
Consider $X=\pres{a,b}{ab=1} \star  \{c\}^*$. Set $\nu(a)=p$,
$\nu(b)=q$, $\nu(c)=1-p-q$ with $p,q>0$, $p+q<1$.  The triple
$(X,\{a,b,c\},\nu)$ is 0-automatic. 

The TE can be solved explicitly. It turns out that there is a unique
solution $\widehat{r}$ which
is determined by:
$$\widehat{r}(a)=\frac{p(1-\widehat{r}(b))}{1-p \widehat{r}(b)} \mbox{ , }
\widehat{r}(b)=\frac{1-\sqrt{1-4pq}}{2p} \mbox{ , }
\widehat{r}(c)=\frac{1-p-q}{1-p \widehat{r}(b)}. $$

According to Theorem \ref{th-rw}, the corresponding drift is
$\widehat{\gamma}=\displaystyle\sqrt{1-4pq}$.
Now, solving the TTE,
we find a unique admissible
solution $(\rho,r)$, given by:
$$\rho=\frac{\lambda(1-p r(b))}{\mu+\lambda p r(b)}, \qquad r =
\widehat{r} \:.
$$

We have $\rho<1$ iff
$\mu>\lambda\widehat{\gamma}=\lambda\displaystyle\sqrt{1-4pq}$.

Let us observe what happens when $\nu(c)$ tends to 0. When
$p<1/2$, there is only one solution $(\rho_1,r_1)$ for the TTE of
the first case. And, as expected, $(\rho,r)$ tends to
$(\rho_1,r_1)$ when $\nu(c)\rightarrow 0$. When $p> 1/2$, there
are two possible solutions $(\rho_1,r_1)$ and $(\rho_2,r_2)$ for
the TTE of the first case. In this  case, $(\rho,r)$ tends to
$(\rho_2,r_2)$ when $\nu(c)\rightarrow 0$.

\begin{figure}[ht]
\[ \epsfxsize=220pt \epsfbox{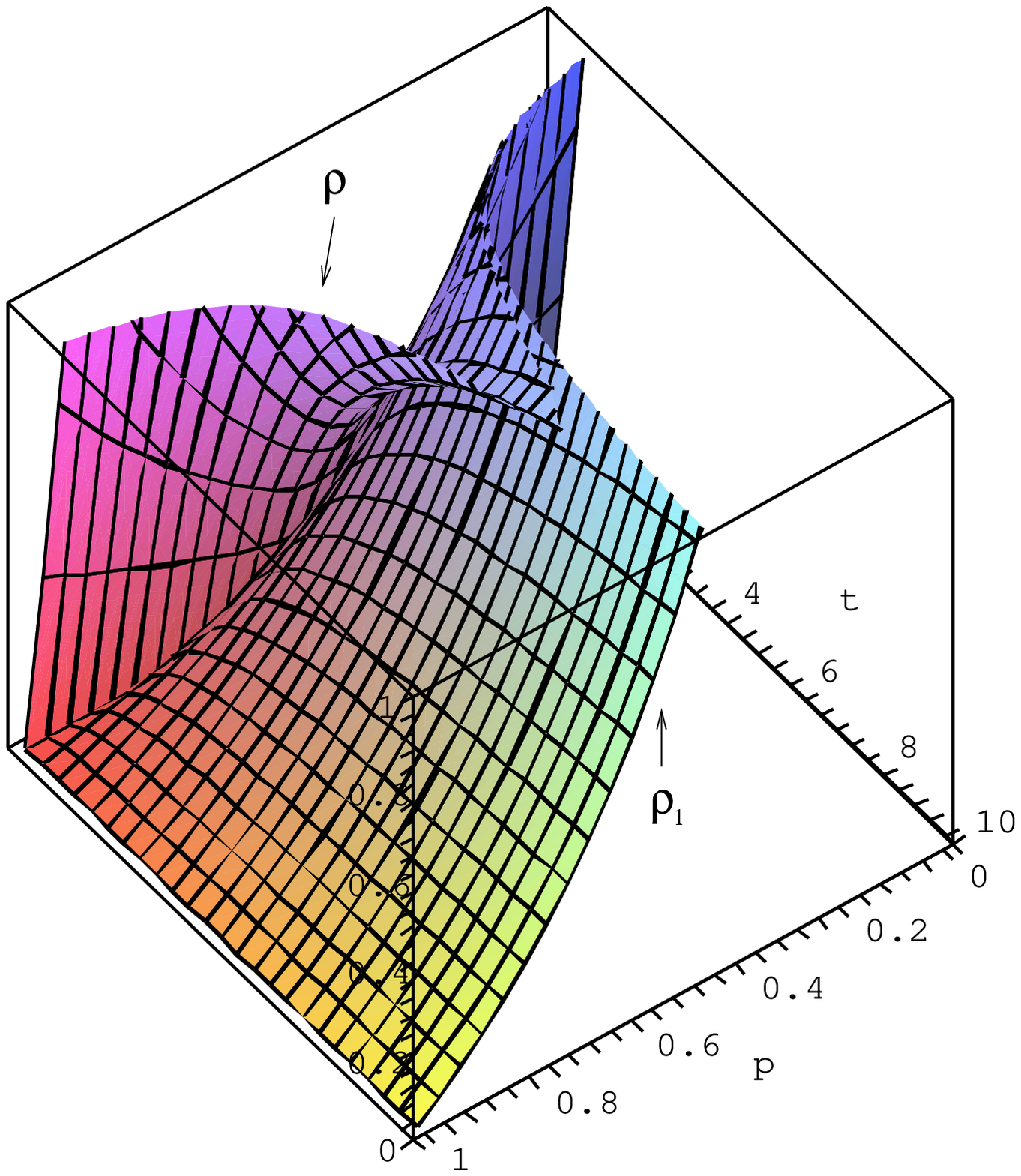}
\epsfxsize=220pt \epsfbox{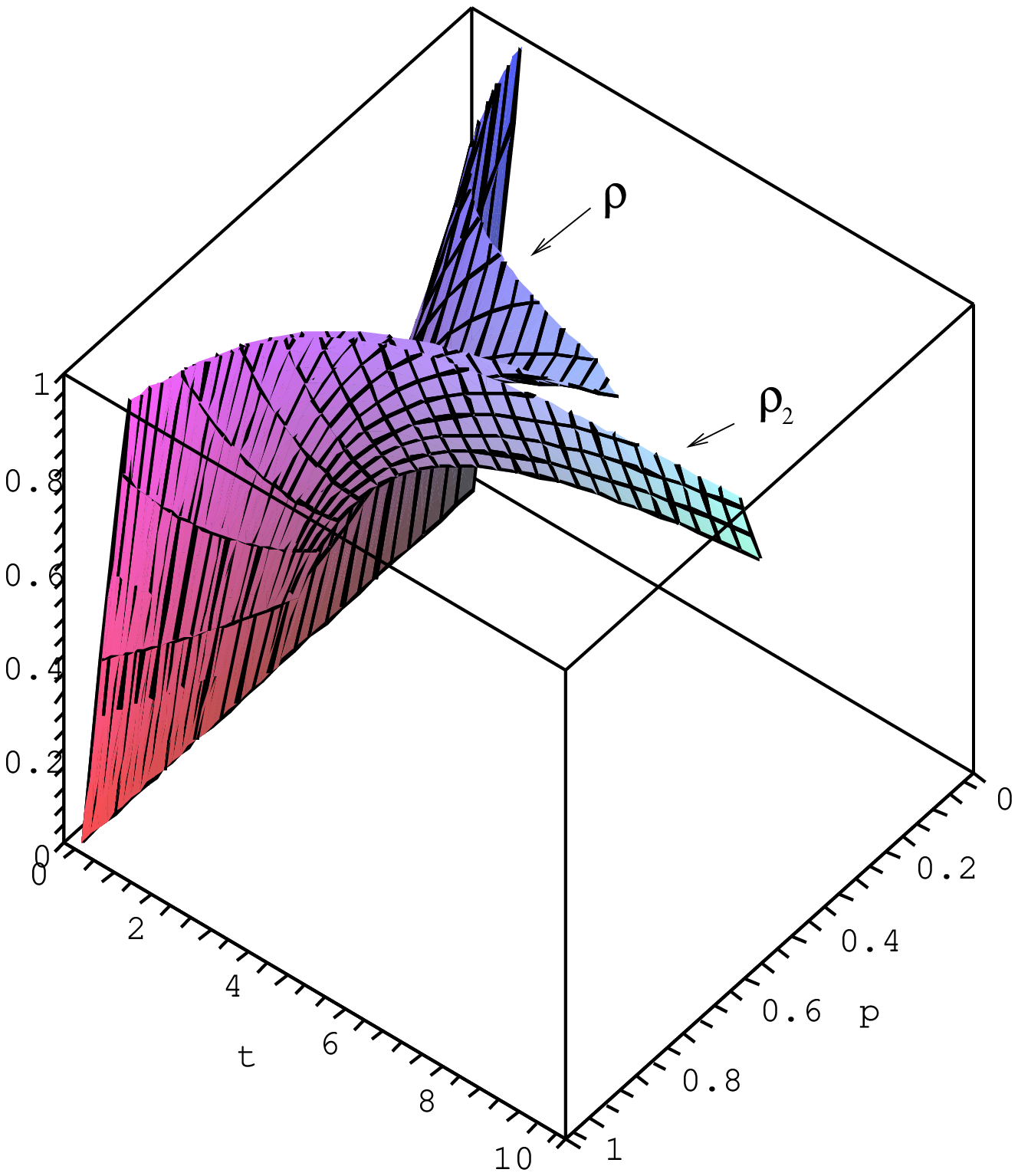} \] 
\caption{$\pres{a,b}{ab=1}$ and $\pres{a,b}{ab=1}\star \{c\}^*$: the
  loads in function of $\nu(a)$ and $t=\lambda/\mu$.}
\label{ab=1_c_p=2/5}
\end{figure}

In Figure \ref{ab=1_c_p=2/5} (left), we
plot $\rho$ and $\rho_1$ as functions of $p$ and $\lambda/\mu$,
for $\nu(c)=0.01$. In Figure \ref{ab=1_c_p=2/5} (right), we
plot $\rho$ and $\rho_2$.

\end{document}

%% file: z2z3.pstex_t
\begin{picture}(0,0)%
\includegraphics{z2z3.pstex}%
\end{picture}%
\setlength{\unitlength}{1184sp}%
\begingroup\makeatletter\ifx\SetFigFont\undefined%
\gdef\SetFigFont#1#2#3#4#5{%
  \reset@font\fontsize{#1}{#2pt}%
  \fontfamily{#3}\fontseries{#4}\fontshape{#5}%
  \selectfont}%
\fi\endgroup%
\begin{picture}(22274,11619)(1439,-11158)
\put(17571,-3045){\makebox(0,0)[lb]{\smash{{\SetFigFont{11}{13.2}{\rmdefault}{\mddefault}{\updefault}$b$}}}}
\put(20362,-2255){\makebox(0,0)[lb]{\smash{{\SetFigFont{11}{13.2}{\rmdefault}{\mddefault}{\updefault}$a$}}}}
\put(23032,-3045){\makebox(0,0)[lb]{\smash{{\SetFigFont{11}{13.2}{\rmdefault}{\mddefault}{\updefault}$b$}}}}
\put(23032,-2255){\makebox(0,0)[lb]{\smash{{\SetFigFont{11}{13.2}{\rmdefault}{\mddefault}{\updefault}$a$}}}}
\put(17514,-5657){\makebox(0,0)[lb]{\smash{{\SetFigFont{11}{13.2}{\rmdefault}{\mddefault}{\updefault}$\nu(a)$}}}}
\put(7868,-7898){\makebox(0,0)[lb]{\smash{{\SetFigFont{11}{13.2}{\rmdefault}{\mddefault}{\updefault}$b^2$}}}}
\put(3901,-1711){\makebox(0,0)[lb]{\smash{{\SetFigFont{11}{13.2}{\rmdefault}{\mddefault}{\updefault}$baba$}}}}
\put(3472,-8814){\makebox(0,0)[lb]{\smash{{\SetFigFont{11}{13.2}{\rmdefault}{\mddefault}{\updefault}$a$}}}}
\put(22198,-5614){\makebox(0,0)[lb]{\smash{{\SetFigFont{11}{13.2}{\rmdefault}{\mddefault}{\updefault}$\nu(b^2)$}}}}
\put(20362,-1588){\makebox(0,0)[lb]{\smash{{\SetFigFont{11}{13.2}{\rmdefault}{\mddefault}{\updefault}$b$}}}}
\put(6879,-4968){\makebox(0,0)[lb]{\smash{{\SetFigFont{11}{13.2}{\rmdefault}{\mddefault}{\updefault}$ba$}}}}
\put(6826,-6511){\makebox(0,0)[lb]{\smash{{\SetFigFont{11}{13.2}{\rmdefault}{\mddefault}{\updefault}$b$}}}}
\put(5026,-8011){\makebox(0,0)[lb]{\smash{{\SetFigFont{11}{13.2}{\rmdefault}{\mddefault}{\updefault}$1$}}}}
\put(9443,-8817){\makebox(0,0)[lb]{\smash{{\SetFigFont{11}{13.2}{\rmdefault}{\mddefault}{\updefault}$b^2a$}}}}
\put(7928,-4006){\makebox(0,0)[lb]{\smash{{\SetFigFont{11}{13.2}{\rmdefault}{\mddefault}{\updefault}$\nu(b^2)$}}}}
\put(23032,-1588){\makebox(0,0)[lb]{\smash{{\SetFigFont{11}{13.2}{\rmdefault}{\mddefault}{\updefault}$b^2$}}}}
\put(20362,-3045){\makebox(0,0)[lb]{\smash{{\SetFigFont{11}{13.2}{\rmdefault}{\mddefault}{\updefault}$b$}}}}
\put(19175,-4923){\makebox(0,0)[lb]{\smash{{\SetFigFont{11}{13.2}{\rmdefault}{\mddefault}{\updefault}$\nu(b)$}}}}
\put(20362,-8929){\makebox(0,0)[lb]{\smash{{\SetFigFont{11}{13.2}{\rmdefault}{\mddefault}{\updefault}$a$}}}}
\put(20362,-9718){\makebox(0,0)[lb]{\smash{{\SetFigFont{11}{13.2}{\rmdefault}{\mddefault}{\updefault}$b$}}}}
\put(4249,-3124){\makebox(0,0)[lb]{\smash{{\SetFigFont{11}{13.2}{\rmdefault}{\mddefault}{\updefault}$bab$}}}}
\put(8071,-3128){\makebox(0,0)[lb]{\smash{{\SetFigFont{11}{13.2}{\rmdefault}{\mddefault}{\updefault}$bab^2$}}}}
\put(4216,-4009){\makebox(0,0)[lb]{\smash{{\SetFigFont{11}{13.2}{\rmdefault}{\mddefault}{\updefault}$\nu(b)$}}}}
\put(4813,-5514){\makebox(0,0)[lb]{\smash{{\SetFigFont{11}{13.2}{\rmdefault}{\mddefault}{\updefault}$\nu(a)$}}}}
\end{picture}%